\UseRawInputEncoding
\documentclass[twocolumn]{aastex631}
\usepackage{amsmath}

\newcommand{\hi}{{\rm H{\sc i} }}

\newcommand{\hii}{{\rm H\,{\sc ii} }}

\def\ltsima{$\; \buildrel < \over \sim \;$}
\def\simlt{\lower.5ex\hbox{\ltsima}}
\def\gtsima{$\; \buildrel > \over \sim \;$}
\def\simgt{\lower.5ex\hbox{\gtsima}}
\def\kms{{\rm\,km\,s^{-1}}}

\def\g{{\rm\,g}}

\def \r {r$^{1/4}$ }


\def\AA{$\; \buildrel \circ \over {\rm A}$}

\def\s{\ifmmode \widetilde \else \~\fi}
\def\={\overline}

\def\spose#1{\hbox to 0pt{#1\hss}}

\def\lta{\mathrel{\spose{\lower 3pt\hbox{$\mathchar"218$}}
     \raise 2.0pt\hbox{$\mathchar"13C$}}}
\def\gta{\mathrel{\spose{\lower 3pt\hbox{$\mathchar"218$}}
     \raise 2.0pt\hbox{$\mathchar"13E$}}}
\def\Dt{\spose{\raise 1.5ex\hbox{\hskip3pt$\mathchar"201$}}}    
\def\dt{\spose{\raise 1.0ex\hbox{\hskip2pt$\mathchar"201$}}}    

\def\r{${\rm r^{1/4}}$}

\def\dotsfill{\leaders\hbox to 1em{\hss.\hss}\hfill}
\def\sun{\odot}
\def\earth{\oplus}

\def\ltsima{$\; \buildrel < \over \sim \;$}
\def\gtsima{$\; \buildrel > \over \sim \;$}
\def\lsim{\lower.5ex\hbox{\ltsima}}
\def\gsim{\lower.5ex\hbox{\gtsima}}
\def\lapp{\ifmmode\stackrel{<}{_{\sim}}\else$\stackrel{<}{_{\sim}}$\fi}
\def\gapp{\ifmmode\stackrel{>}{_{\sim}}\else$\stackrel{<}{_{\sim}}$\fi}

\submitjournal{ApJ}

\shorttitle{Young, blue, and isolated stellar systems in the Virgo Cluster\\
I. 2-D Optical spectroscopy}
\shortauthors{Bellazzini et al.}
\graphicspath{{./}{figures/}}

\begin{document}

\title{Young, blue, and isolated stellar systems in the Virgo Cluster\\
I. 2-D Optical spectroscopy}

\correspondingauthor{Michele Bellazzini}
\email{michele.bellazzini@inaf.it}

\author[0000-0001-8200-810X]{Michele Bellazzini}
\affil{INAF â Osservatorio di Astrofisica e Scienza dello Spazio di Bologna, Via Gobetti 93/3, 40129 Bologna, Italy}

\author[0000-0003-4486-6802]{Laura Magrini}
\affil{INAF - Osservatorio Astrofisico di Arcetri, Largo E. Fermi 5, 50125 Firenze, Italy}

\author[0000-0002-5434-4904]{Michael G. Jones}
\affiliation{Steward Observatory, University of Arizona, 933 North Cherry Avenue, Rm. N204, Tucson, AZ 85721-0065, USA}

\author[0000-0003-4102-380X]{David J. Sand}
\affiliation{Steward Observatory, University of Arizona, 933 North Cherry Avenue, Rm. N204, Tucson, AZ 85721-0065, USA}

\author[0000-0002-3865-9906]{Giacomo Beccari}
\affil{European Southern Observatory, Karl-Schwarzschild-Stra{\ss}e 2, D-85748 Garching bei M\"{u}nchen, Germany}

\author[0000-0002-5281-1417]{Giovanni Cresci}
\affil{INAF - Osservatorio Astrofisico di Arcetri, Largo E. Fermi 5, 50125 Firenze, Italy}

\author[0000-0002-0956-7949]{Kristine Spekkens}
\affiliation{Department of Physics and Space Science, Royal Military College of Canada P.O. Box 17000, Station Forces Kingston, ON K7K 7B4, Canada}
\affiliation{Department of Physics, Engineering Physics and Astronomy, Queens University, Kingston, ON K7L 3N6, Canada}

\author[0000-0001-8855-3635]{Ananthan Karunakaran}
\affiliation{Instituto de Astrof\'{i}sica de Andaluc\'{i}a (CSIC), Glorieta de la Astronom\'{i}a, 18008 Granada, Spain}
\affiliation{Department of Physics, Engineering Physics and Astronomy, Queen’s University, Kingston, ON K7L 3N6, Canada}

\author[0000-0002-9798-5111]{Elizabeth A. K. Adams}
\affil{ASTRON, Netherlands Institute for Radio Astronomy, Oude Hoogeveensedijk 4, 7991 PD Dwingeloo, The Netherlands}
\affil{Kapteyn Astronomical Institute, University of Groningen, PO Box 800, 9700 AV Groningen, The Netherlands}

\author[0000-0002-5177-727X]{Dennis Zaritsky}
\affiliation{Steward Observatory, University of Arizona, 933 North Cherry Avenue, Rm. N204, Tucson, AZ 85721-0065, USA}

\author[0000-0002-6551-4294]{Giuseppina Battaglia}
\affil{Instituto de Astrof\'isica de Canarias, V\'ia L\'actea s/n 38205 La Laguna, Spain}
\affil{Department of Astrophysics, University of La Laguna,San Crist{\'o}bal de La Laguna,E-38206, Spain}

\author[0000-0003-0248-5470]{Anil Seth}
\affiliation{Department of Physics \& Astronomy, University of Utah, Salt Lake City, UT, 84112, USA}

\author[0000-0002-1821-7019]{John M. Cannon}
\affiliation{Department of Physics \& Astronomy, Macalester College, 1600 Grand Avenue, Saint Paul, MN 55105, USA}

\author[0000-0002-8598-439X]{Jackson Fuson}
\affiliation{Department of Physics \& Astronomy, Macalester College, 1600 Grand Avenue, Saint Paul, MN 55105, USA}

\author[0000-0002-9724-8998]{John L. Inoue}
\affiliation{Department of Physics \& Astronomy, Macalester College, 1600 Grand Avenue, Saint Paul, MN 55105, USA}

\author[0000-0001-9649-4815]{Bur\c{c}in Mutlu-Pakdil}
\affil{Kavli Institute for Cosmological Physics, University of Chicago, Chicago, IL 60637, USA}
\affil{Department of Astronomy and Astrophysics, University of Chicago, Chicago IL 60637, USA}

\author[0000-0001-8867-4234]{Puragra Guhathakurta}
\affiliation{UCO/Lick Observatory, University of California Santa Cruz, 1156 High Street, Santa Cruz, CA 95064, USA}

\author[0000-0002-0810-5558]{Ricardo R. Mu\~{n}oz}
\affiliation{Departamento de Astronom\'ia, Universidad de Chile, Camino El Observatorio 1515, Las Condes, Santiago}

\author[0000-0001-8354-7279]{Paul Bennet}
\affiliation{Space Telescope Science Institute, 3700 San Martin Drive, Baltimore, MD 21218, USA}

\author[0000-0002-1763-4128]{Denija Crnojevi\'{c}}
\affil{University of Tampa, 401 West Kennedy Boulevard, Tampa, FL 33606, USA}

\author[0000-0003-2352-3202]{Nelson Caldwell}
\affiliation{Center for Astrophysics, Harvard \& Smithsonian, 60 Garden Street, Cambridge, MA 02138, USA}

\author[0000-0002-1468-9668]{Jay Strader}
\affiliation{Center for Data Intensive and Time Domain Astronomy, Department of Physics and Astronomy, Michigan State University, East Lansing, MI 48824, USA}

\author[0000-0001-6443-5570]{Elisa Toloba}
\affiliation{Department of Physics, University of the Pacific, 3601 Pacific Avenue, Stockton, CA 95211, USA}

\begin{abstract}
We use panoramic optical spectroscopy obtained with MUSE@VLT to investigate the nature of five candidate extremely isolated low-mass star forming regions (Blue Candidates, BCs hereafter) toward the Virgo cluster of galaxies. Four of the five (BC1, BC3, BC4, BC5) are found to host several \hii regions and to have radial velocities fully compatible with being part of the Virgo cluster. All the confirmed candidates have mean metallicity significantly in excess of that expected from their stellar mass, indicating that they originated from gas stripped from larger galaxies. In summary, these four candidates share the properties of the prototype system SECCO~1, suggesting the possible emergence of a new class of stellar systems, intimately linked to the complex duty cycle of gas within clusters of galaxies. A thorough discussion on the nature and evolution of these objects is presented in a companion paper, where the results obtained here from MUSE data are complemented with {\em Hubble Space Telescope} (optical) and {\em Very Large Array} (H{\sc i}) observations.
\end{abstract}


\section{Introduction}
\label{sec:intro}
The publication of catalogues of compact \hi sources from the ALFALFA \citep{Adams2013} and GALFA \citep{Saul2012} surveys triggered observational campaigns aimed at detecting their stellar counterparts, in search of new very dark local dwarf galaxies hypothesised to be associated with the gas clouds \citep{sand15,secco_surv,tollerud2015}. The new experiments failed to find a significant population of such objects \citep[see, e.g.,][]{becca16}, still a few new interesting stellar systems were identified \citep[see, e.g.,][]{LeoP2013,LeoP2015,tollerud2015,cannon15,bennet2022}.

One of the most curious cases is SECCO~1 \citep[also known as AGC 226067;][]{secco_surv,secco1_15,sand15,Adams2015}, a low-mass ($M_{\star}\simeq 10^5~M_{\sun}$, $M_{HI}\simeq 2\times 10^7~M_{\sun}$) star-forming stellar system lying within the Virgo cluster of galaxies. Given the very low stellar mass, the high mean metallicity ($\langle 12+{\rm log}(O/H)\rangle = 8.38\pm 0.11$) implies that the gas fuelling star formation in SECCO~1 was stripped from a relatively massive gas-rich galaxy \citep[][]{secco1_17,sand17,secco1_18}, either by a tidal interaction or by ram pressure exerted by the hot intra cluster medium (ICM). Indeed, star formation is known to occur in gas clouds stripped from galaxies via both channels \citep[see, e.g.,][and references therein]{pasha2021,poggianti2019}. However, in both cases, the star-forming stripped knots are always seen in proximity to the parent galaxy and/or connected to the parent galaxy either by tidal tails or by the jellyfish structures that are the classical fingerprint of ram-pressure stripping \citep[see, e.g.,][]{Gerhard2002,yoshida2012,fuma2011,fuma2014,kenney2014,fossati2016, mb_pw1,nide_PW1,corb2021a,corb2021b}. In contrast, SECCO~1 is extremely isolated, lying more than 200~kpc, in projection, from the nearest candidate parent galaxy \citep{sand17,secco1_18,jones_ALLBC}.

Simple theoretical arguments \citep{burk2016} as well as dedicated hydrodynamical simulations \citep{secco1_18,Calura20}, suggest that gas clouds similar to SECCO~1 may survive as long as $\sim 1$~Gyr within clusters of galaxies, kept together by the pressure confinement of the ICM, thus leaving room for very long voyages from the site of origin. Star formation is also expected to occur in the meanwhile \citep[][]{Calura20,kapfe}. If cloudlets such as these are indeed able to survive such long times and form stars, a rich population of them should be among the inhabitants of galaxy clusters, given the complex processes in which the gas is involved in these environments \citep{poggianti2019,ram_rev2021}. 

Prompted by these considerations, a search for similar systems in the Virgo cluster was performed \citep[see][]{sand17,jones_BC3} and a sample of five Blue Candidates (BCs) were selected from UV and optical images, as described in \citet{jones_ALLBC}. To confirm the nature of these BCs, spectroscopic follow-up is essential to a) measure their radial velocity and confirm their location within the Virgo Cluster; b) quantify any ongoing star formation by detecting and analyzing associated \hii\, regions, and c) estimate the metallicity of the gas in star forming regions, a crucial diagnostic for assessing the origin of these systems.

Here we present the results of VLT@MUSE \citep{muse} observations of five such blue candidate stellar systems, fully analogous to the study by \citet[][Be17 hereafter]{secco1_17} for SECCO~1. Four of them are confirmed as likely residing in Virgo and similar to SECCO~1. Individual \hii regions are identified, velocities and line fluxes are measured from their spectra and metallicity estimates are provided. Finally, their chemical and kinematic properties are briefly discussed. In a companion paper \citep[paper~II;][Pap-II hereafter]{jones_ALLBC}, {\em Hubble Space Telescope} ({\em HST}) imaging and photometry as well as new {\em Very Large Array (VLA)} and {\em Green Bank telescope (GBT)} \hi observations for the BCs are presented. 
The companion paper also discusses the nature and origin of this potentially new class of stellar systems, based on the entire set of available observations. 
A more detailed discussion on BC3, also known as AGC~226178, the only system that has been subject of previous analyses \citep[][]{cannon15,junais21}, has been presented by \citet{jones_BC3}.

\begin{figure*}[!htbp]
\gridline{\fig{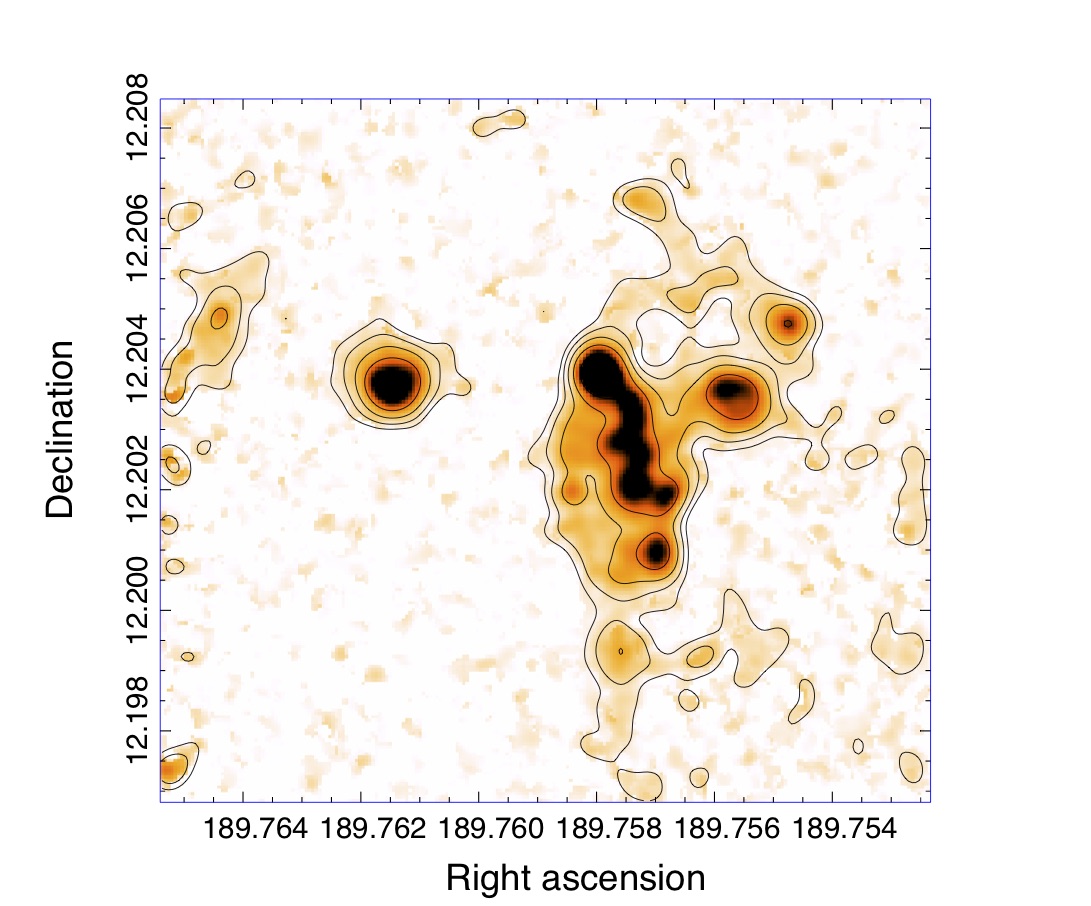}{0.5\textwidth}{(a)}
          \fig{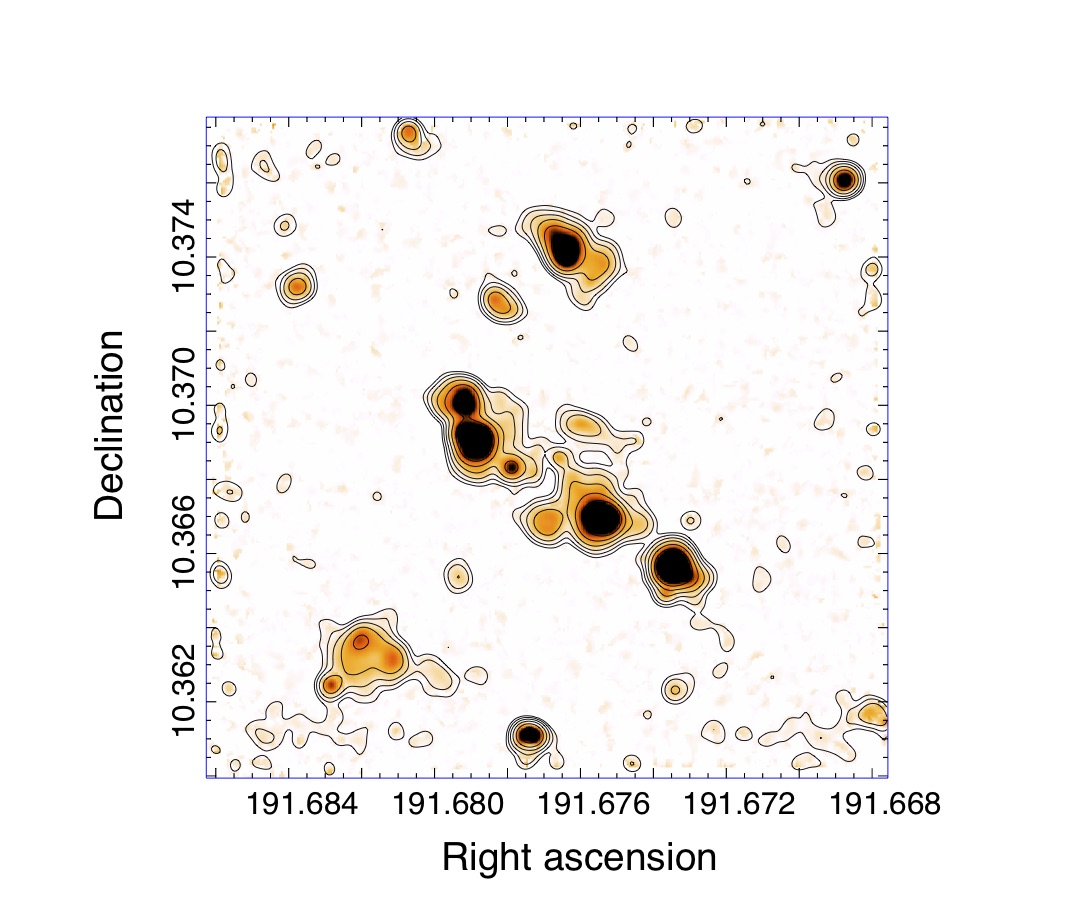}{0.5\textwidth}{(b)}
          }
\caption{Maps of the H$\alpha$ emission from BC1 (a) and BC3 (b). The levels of the contours are at 2, 4, 8, 16, 32, 64, 128, 256 $\times 10^{-20}$ erg cm $^{-2}$ s$^{-1}$. { Right ascension and declination are in degrees.}
\label{fig:map13}}
\end{figure*}

\begin{figure*}[!htbp]         
\gridline{\fig{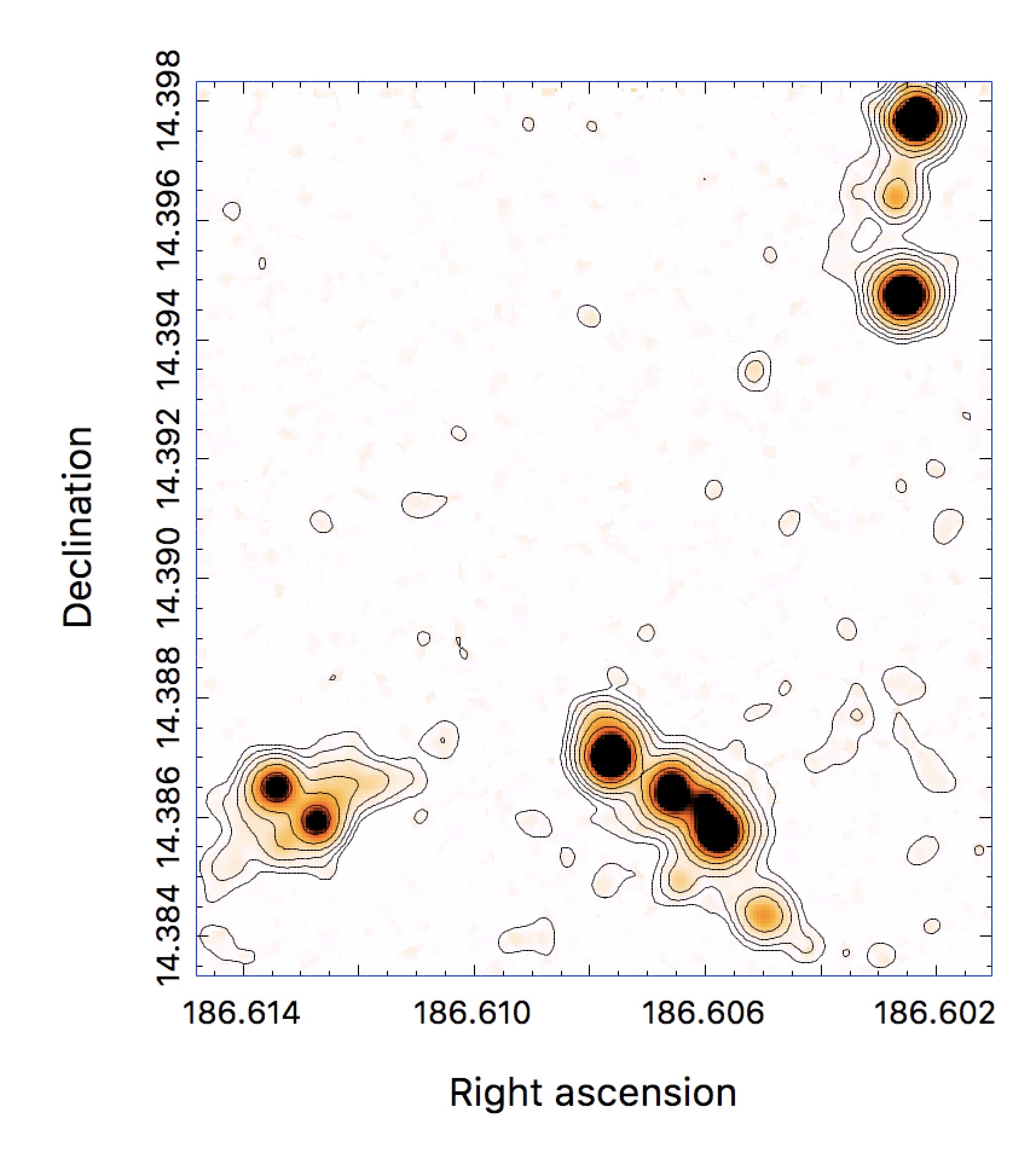}{0.4\textwidth}{(a)}
          \fig{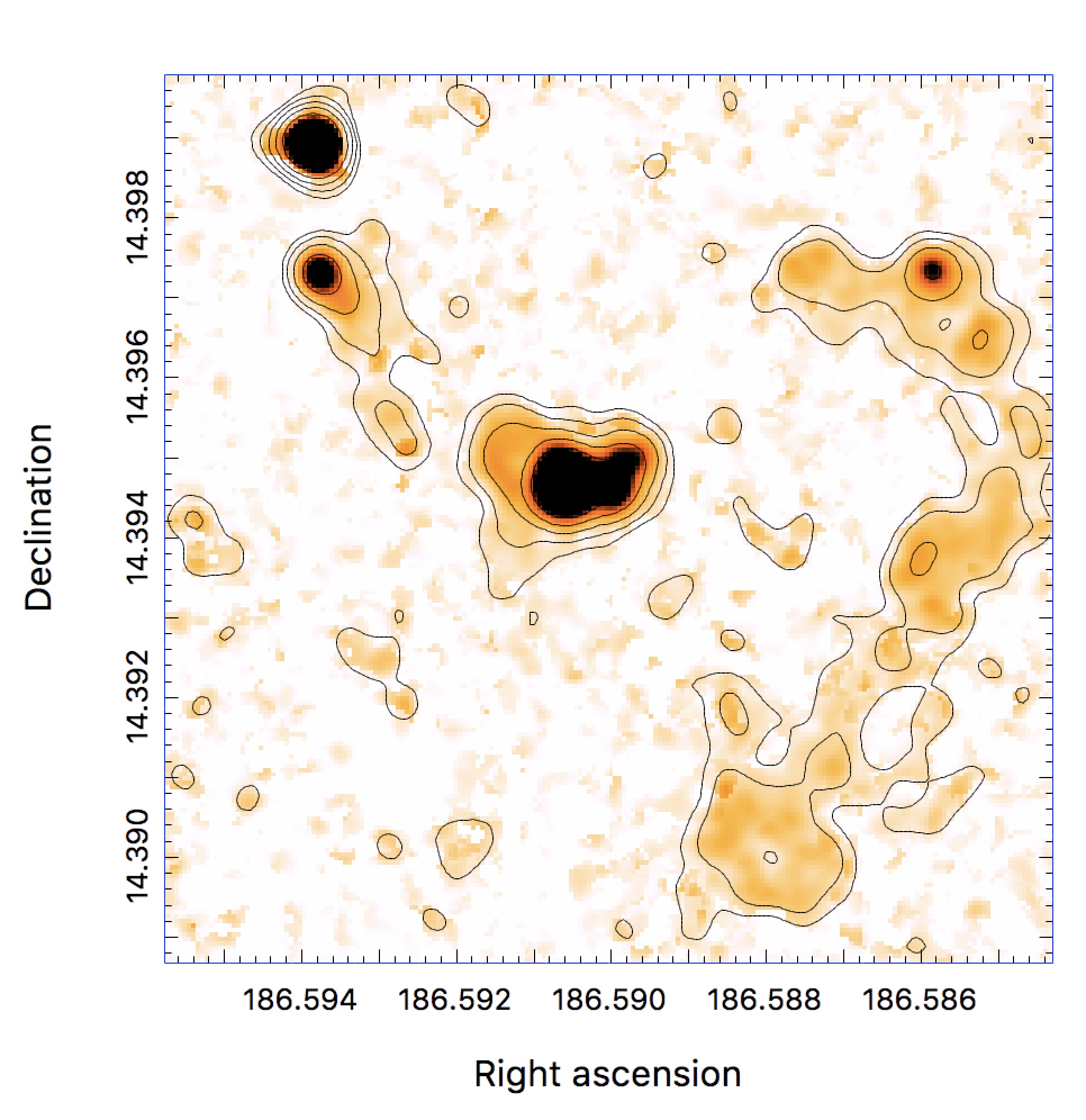}{0.43\textwidth}{(b)}
          }
\gridline{\fig{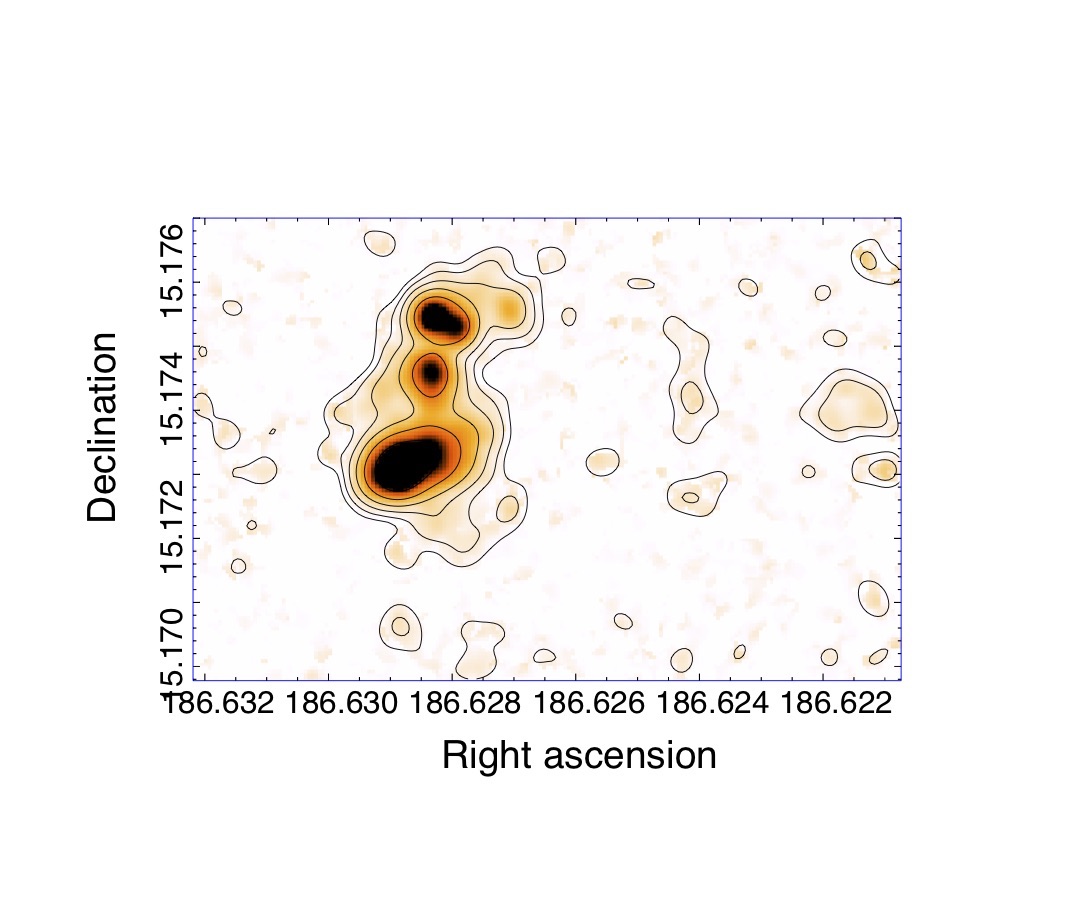}{0.43\textwidth}{(c)}}
\caption{Maps of the H$\alpha$ emission from BC4L (a), BC4R (b), and BC5 (c). The levels of the contours are at 2, 4, 8, 16, 32, 64, 128, 256 $\times 10^{-20}$ erg cm $^{-2}$ s$^{-1}$. { Right ascension and declination are in degrees.}
\label{fig:map45}}
\end{figure*}

\section{Observations and data reduction}
\label{sec:obse}

Integral field unit optical (4650-9300\AA) spectroscopy of six $1.0\arcmin \times 1.0\arcmin$ fields centered on the targets was acquired with the Multi Unit Spectroscopic Explorer \citep[MUSE;][]{muse}, mounted at the Unit~4 (Yepun) Very Large Telecope (VLT), at ESO, Paranal (Chile), as part of program 0101.B-0376A (P.I: R. Mun\~oz). The spectral resolution is in the range $\frac{\lambda}{\Delta \lambda}\simeq 2000 - 4000$, from the bluest to the reddest wavelength.
Two partially overlapping MUSE pointings were required to sample all the sources presumably associated with BC4. We refer to these two fields and portions of the BC4 system as BC4L (left) and BC4R (right), BC4L lying to the East-South-East of BC4R. For each field, six $t_{exp}=966$~s exposures were acquired with a dithering scheme based on regular de-rotator offsets to improve flat-fielding and homogeneity of the image quality across the field. The observing log is presented in Table~\ref{tab:targets}. Each set of raw data were wavelength and flux calibrated, and then stacked into a single, final data-cube per field using the MUSE pipeline \citep{weilbacher2012}.

The spectra of the sources associated with each candidate were visually inspected using SAOimage DS9, looking for H$\alpha$ emission in the redshift range compatible with membership to the Virgo cluster \citep[$-500~\kms \la cz \la 3000~\kms$;][]{mei2007}. This was easily found in all the sources except BC2. As discussed in Pap-II, BC2 appears to be a small group of background blue galaxies in the {\it HST} imaging, mimicking the appearance of the other BC objects in our sample.
This, combined with the lack of H$\alpha$ emission, led us to dismiss it as a spurious candidate and not analyse it further.
As in Pap-II, we adopt for all our targets the distance to the Virgo cluster by \citet{mei2007}, $D=16.5$~Mpc.

\begin{table}
\caption{MUSE fields}
\label{tab:targets}
\hskip -2cm
\begin{tabular}{lccccc}
\hline
\hline
name&RA$_{J2000}$ & Dec$_{J2000}$ & Date Obs.  &  t$_{exp}$	 &    FWHM$^1$ \\
BC1 & 189.756116  &  12.20542	  & 2018-05-17 & 966 s$\times$ 6 &	$0.3\arcsec$	\\
BC2 & 191.114323  &  12.61824	  & 2018-05-19 & 966 s$\times$ 6 &	$0.5\arcsec$	\\
BC3 & 191.677299  &  10.36919	  & 2019-02-27 & 966 s$\times$ 6 &	$1.1\arcsec$	\\
BC4L& 186.608125  &  14.38914	  & 2019-02-28 & 966 s$\times$ 6 &	$0.8\arcsec$	\\
BC4R& 186.591389  &  14.39417	  & 2019-02-28 & 966 s$\times$ 6 &	$0.5\arcsec$	\\
BC5 & 186.63014   &  15.1745	  & 2019-02-10 & 966 s$\times$ 6 &	$0.5\arcsec$	\\
\hline
\hline
\end{tabular}
\tablecomments{$^1$ Full Width at Half Maximum of the seeing as recorded into the header of the data-cubes.}
\end{table}


For the detection of the individual sources within each field and to extract their spectra we adopted
the same approach as Be17. In particular, for each stacked data-cube we proceeded as follows:

\begin{enumerate}  

\item The data-cube was split into 3801 single layers, sampling the targets from 4600.29\AA~ to 9350.29\AA, with a step in wavelength of 1.25\AA.

\item An H$\alpha$ image was produced by stacking together the four layers where the local H$\alpha$ signal reached its maximum, corresponding to a spectral window of 5.0\AA, and a white image was produced stacking together all 3801 layers.

\item Both images were searched for sources with intensity peaks above $3.0\sigma$ from the background level using the photometry and image analysis package {\tt Sextractor} \citep{sextra}. The two lists of detected sources where then merged together into a single master list.

\item Photometry through an aperture of radius $1.5\arcsec$ was performed with {\tt Sextractor} on each individual layer image for all the sources included in the master list.

\item The fluxes measured in each layer were then recombined, obtaining a 1D spectrum for each source.

\item Finally, the spectrum of all the measured sources was visually inspected and only those having 
clear emission at least in H$\alpha$ were retained in the final catalogue, which is presented in Table~\ref{tab:RVpos}. Sources that were originally detected only in the white light image are denoted by a W at the end of their names. 

\end{enumerate}

We identify emission lines across the MUSE spectral range including H$\beta$, [O {\sc iii}]4959, [O {\sc iii}]5007, [N {\sc ii}]6548, H$\alpha$, [N {\sc ii}]6583, [S {\sc ii}]6717, and [S {\sc ii}]6731.
The heliocentric radial velocity (RV) of each of the 53 sources included in the final catalogue has been measured by
fitting all the identified emission lines with a Gaussian curve to estimate their centroid and, consequently, the shift with respect to their rest wavelength with the {\sc IRAF} task {\sc rvidlines}. The final RV was derived from the average wavelength shift, while the uncertainty ($\epsilon{RV}$) is the associated rms divided by the square root of the number of lines involved in the estimate ($N_{RV}$). Based on the scatter between the velocities of different lines for the same source, an uncertainty of 20.0$\kms$ was adopted for the sources whose RV has been estimated from one single line (H$\alpha$). { Position, radial velocity (and associated uncertainty), H$\alpha$ flux (and associated uncertainty), and the Full Width at Half Maximum (FWHM) of all the identified sources are listed in Table~\ref{tab:RVpos}}.

It is possible that the spectra of some of the sources detected in white light are the combination of an emission component (coming from the diffuse hot gas associated with the BC), superimposed on an unrelated background/foreground source (whose flux triggered the detection in white light).  
We decided to keep these sources in the final list as, in any case, they provide additional sampling of the velocity fields and of the oxygen abundance of the considered systems. Several of the sources listed in Tab.~\ref{tab:RVpos} have FWHM, as measured by {\tt Sextractor}, significantly larger than the seeing (see Tab.~\ref{tab:targets}), indicating that they are extended. In some cases the FWHM is also significantly larger than the aperture we used to extract the spectra from the data-cube. We found that the adopted aperture radius of $1.5\arcsec$ is a reasonable trade-off to maximise the signal-to-noise ratio of the average source while minimising the contamination from adjacent sources.

Line fluxes and the associated uncertainties have been obtained with the task {\sc splot} in {\sc IRAF}, as done in Be17, and are listed in Table~\ref{tab:lineFlux}, for the subset of 37 sources having valid measures of both the H$\alpha$ and H$\beta$ fluxes. These allow an estimate of  
the extinction ($C_{\beta}$, also listed in Tab.~\ref{tab:lineFlux}), computed from the ratio between the observed and theoretical Balmer decrements for the typical conditions of an \hii region \citep[see][]{oster2006}. 
 Within the same complex of \hii\,  regions there can be significant reddening differences. The measured extinction is both that within each \hii\, region and that outside, between the observer and the object \citep[see, e.g,][]{caplan86}. Some regions have likely higher internal extinction than others in the same group, such as BC1s11 and BC3s18, the most extincted regions of our sample. Also, within each BC we notice a variation in the degree of ionisation among the various \hii\,  regions, evidenced by the observation of [O{\sc iii}] lines in only some of them. These variations are very similar to those observed in SECCO~1 (Be17) and, as in that case, they do not appear to be associated to metallicity variations, but to the temperature of the ionising stars \citep[see, e.g.,][]{secco1_18}.

Uncertainties in line fluxes can be very large in some cases, in particular for [N{\sc ii}] and [S {\sc ii}] lines, owing to low signal to noise. Still we preferred to keep these measures as they may bring useful information to derive average properties of the stellar systems.

\section{Morphology, Classification and Kinematics}
\label{sec:kine}

In Fig.~\ref{fig:map13} and Fig.~\ref{fig:map45} we show the continuum-subtracted H$\alpha$ images of BC1, BC3, BC4L, BC4R and BC5, with intensity contours ranging from $2\times 10^{-20}$ erg cm $^{-2}$ s$^{-1}$ to $256 \times 10^{-20}$ erg cm $^{-2}$ s$^{-1}$, spaced by a factor of 2. 
Each image was obtained by subtracting a continuum image from the H$\alpha$ image described above. The continuum image was created by stacking four slices of the cube near the emission line, in particular from 2.5\AA~ to 7.5\AA~ blueward of the analogous window centered on H$\alpha$. { In other words, the continuum image has the same wavelength width as the H$\alpha$ image but is shifted by $\simeq 7.5$\AA~ to the blue of the 
H$\alpha$ line.}

All the images display objects whose morphology is very similar to SECCO~1: several compact sources, often distributed in elongated configurations, which are surrounded by diffuse ionised emission. All the systems, except BC5\footnote{In fact there is one source, BC5s3, that is only partially imaged by the BC5 data-cube, lying at its southern edge. It is located $\simeq 0.7\arcmin$ apart from the main BC5 clump of sources shown in Fig.~\ref{fig:map45}c. A map of BC5 sources including BC5s3 is presented in Sect.~\ref{sec:kine}.}, appear fragmented into separate pieces, with typical separation 
$\la 0.5\arcmin$, corresponding to $\la 2.4$~kpc at the distance of Virgo. The two pieces of BC4 are separated by 
$\sim 8$~kpc, in projection, very similar to the separation between the Main Body and the Secondary Body of SECCO~1,  \citep[MB and SB;][]{sand15,secco1_18}. The extension of the different systems is different by a factor of a few. The bright H$\alpha$ knots of BC1  can be approximately enclosed within a circle of projected radius $\simeq 1.4$~kpc; the same radius for BC3, BC4L, BC4R, and BC5 is $\simeq 1.8, 2.2, 1.2$ and $0.5$~kpc, respectively. A deeper insight into the morphology of the BCs is presented in Pap-II, based on the inspection of {\em HST} images.

\begin{figure}[!htbp]
\includegraphics[width=\columnwidth]{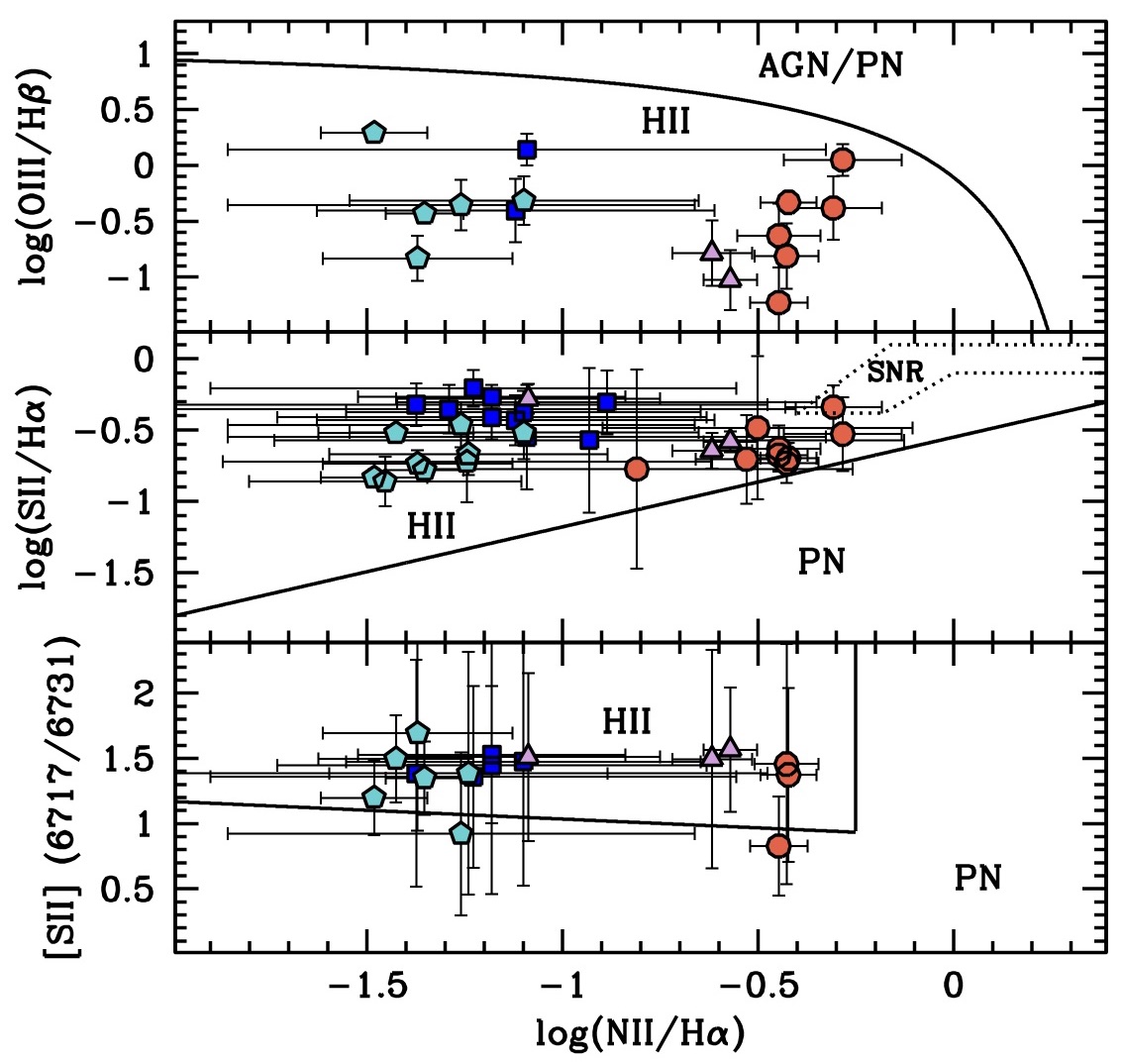}
\caption{Line-ratio diagnostic diagrams for all the individual sources in which it was possible to measure the relevant line fluxes from MUSE spectra. The symbol coding is as follows: blue squares BC1, turquoise pentagons BC3, red circles BC4, plum triangles BC5. The lines separating different kind of sources in the diagrams are: Eq. 5 from Kewley et al. 2001 (upper panel), Eq. 3 from Kniazev et al. 2008 (middle panel), and Eq. 4 and 5 from the same source (lower panel). { In the middle panel, a dotted line contour encloses the region of the diagram where Supernova Remnants (SNR) are expected to lie}.}
\label{fig:diagno}
\end{figure}

\begin{figure}[!htbp]
\includegraphics[width=\columnwidth]{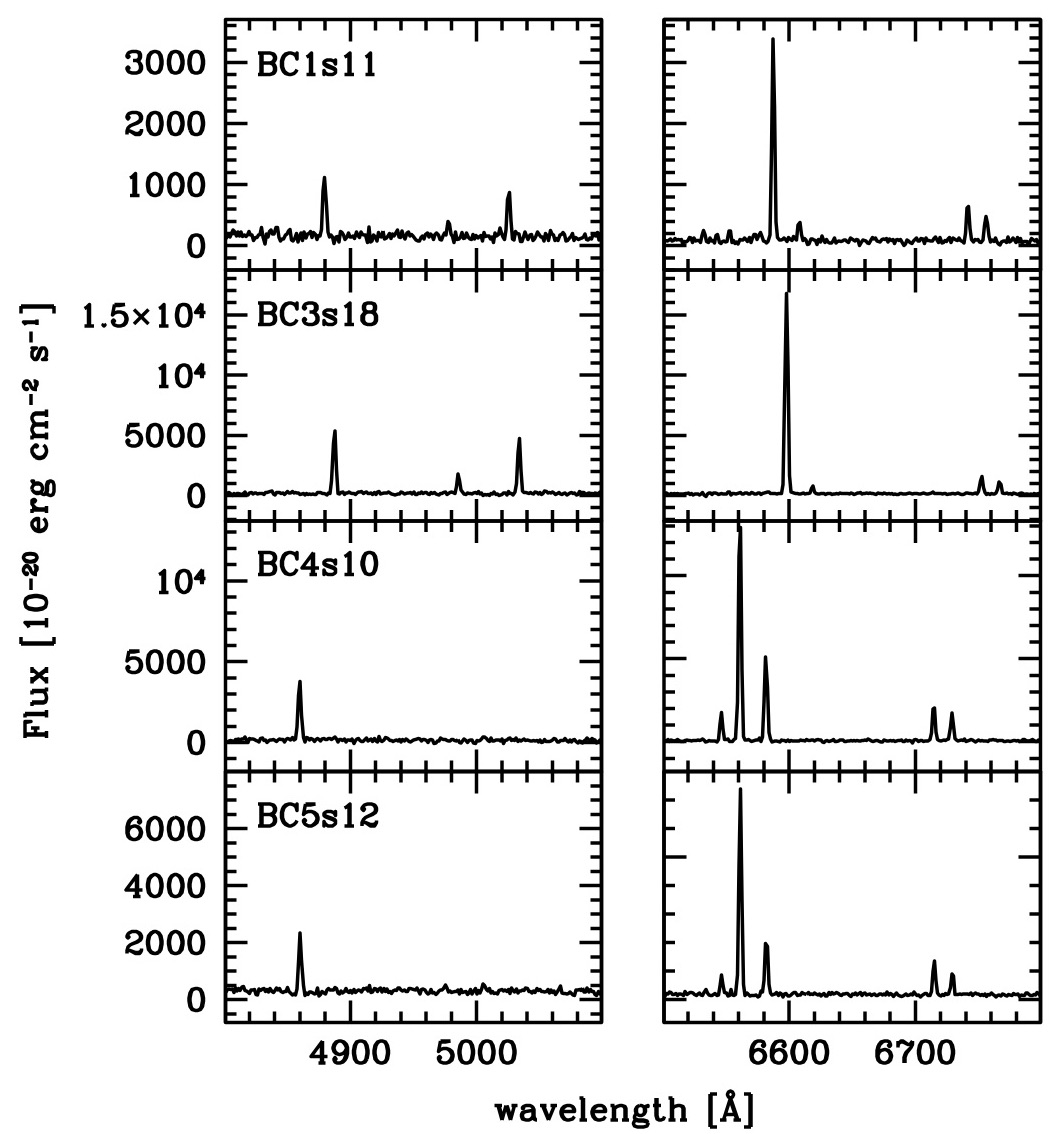}
\caption{Spectra of the sources with the strongest H$\alpha$ lines in each BC. For each source, two portions of the MUSE spectra are shown, enclosing the most relevant emission lines: H$\beta$, [O {\sc iii}]4959, [O {\sc iii}]5007 (left panels), and [N {\sc ii}]6548, H$\alpha$, [N {\sc ii}]6583, and [S {\sc ii}]6717$+$6731 (right panels).}
\label{fig:spe}
\end{figure}

The issue of the classification of the individual sources detected with {\tt Sextractor} is addressed in Fig.~\ref{fig:diagno}, where three different diagnostic plots based on line ratios are presented for the (different) subsets of sources having measures of the flux in the involved lines. While the uncertainties are large in some cases, all the considered sources behave as \hii regions, { the only possible exception being BC2s12L that lies just within the contour enclosing Supernova remnants in the middle panel of Fig.~\ref{fig:diagno}}. We conclude that all the systems are actively forming stars, fully analogous to the case of SECCO~1. In Fig.~\ref{fig:spe} we show the spectra of the sources with the strongest H$\alpha$ line in each BC, to show the quality of the best spectra in our dataset.

\begin{figure*}[!htbp]
\gridline{\fig{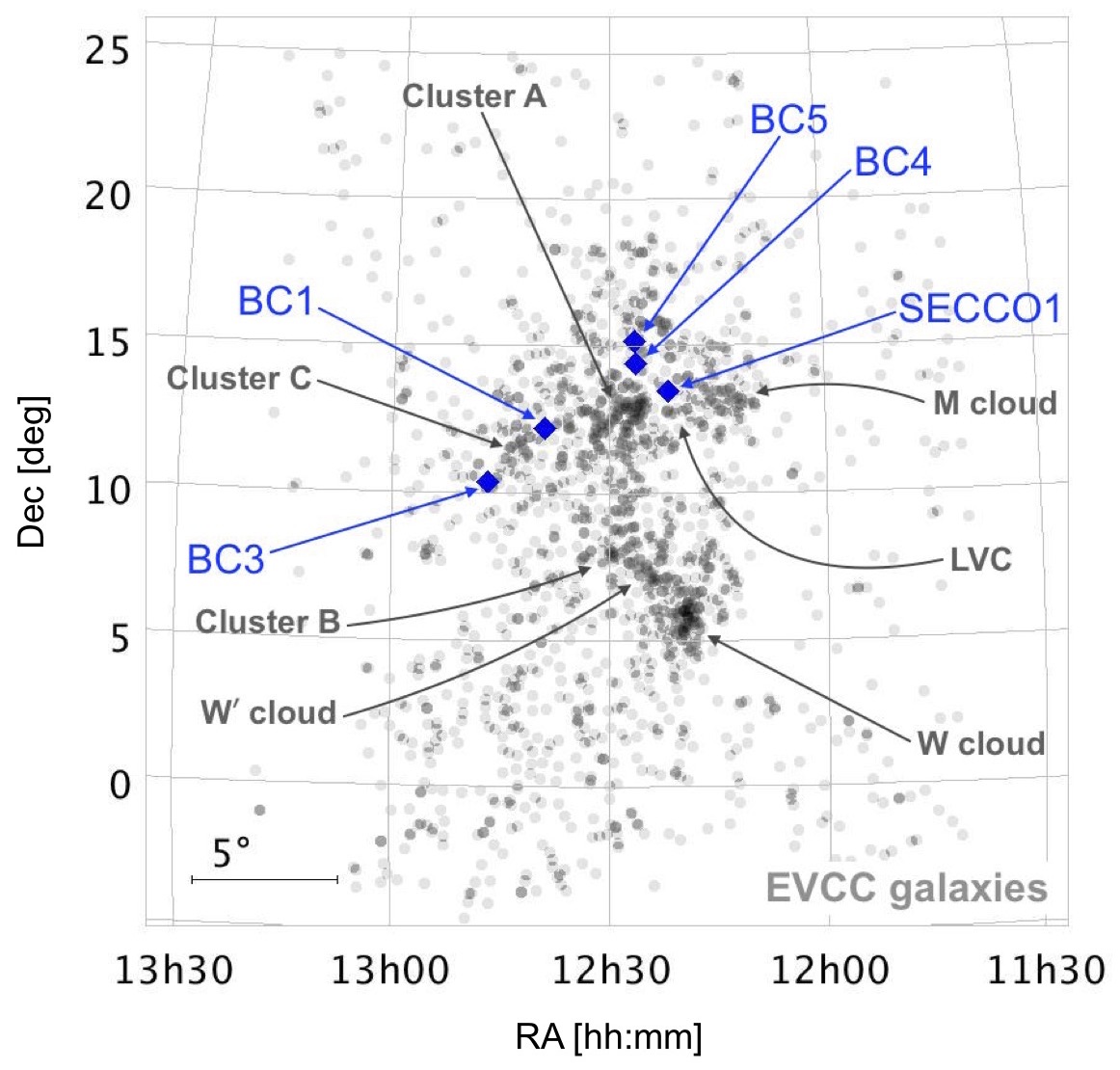}{0.48\textwidth}{(a)}
          \fig{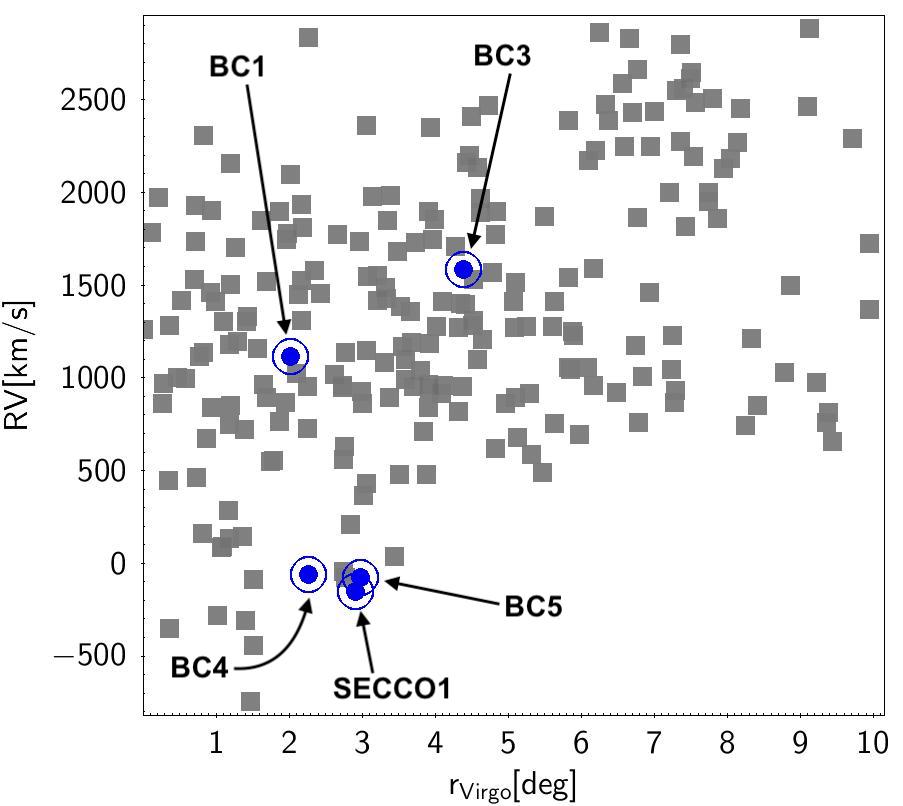}{0.5\textwidth}{(b)}
         }
\caption{Location of the Blue Candidates and of SECCO~1 within the Virgo cluster. In panel (a) the position of the systems (blue diamonds) is indicated within a wide map of the Virgo cluster, as traced by the distribution of EVCC galaxies \citep[small grey circles;][]{EVCC}; the main substructures of the cluster are labelled, following \citep{boselli14}. In panel (b) EVCC galaxies (grey squares) and the BCs 
(encircled blue filled circles) are plotted into a phase-space diagram opposing the heliocentric line of sight velocity to the angular distance from M~87, taken as the center of the Virgo cluster. 
\label{fig:virgomap}}
\end{figure*}

Before proceeding with the analysis of the internal kinematics of the new stellar systems it may be worth putting them
in context within the Virgo cluster of galaxies. In Fig.~\ref{fig:virgomap}a, BC1, BC3, BC4, and BC5, together with SECCO~1, are shown in projection on a wide map of Virgo, as traced by the distribution of galaxies included in the Extended Virgo Cluster Catalog \citep[EVCC;][]{EVCC}. The main substructures of the cluster are labelled, following \cite{boselli14}. Fig.~\ref{fig:virgomap}b shows that all our targets have mean velocities (from Tab.~\ref{tab:means}) within the range spanned by EVCC galaxies, hence they are very likely members of the Virgo cluster, a conclusion that is also supported by {\em HST} data (Pap-II)\footnote{ For example, by resolving the BCs into stars and showing that their color magnitude diagrams are consistent with young stellar population at the distance of Virgo \citep[see also][]{jones_BC3}}. In particular, BC1 and BC3 are consistent with membership to Cluster C, while SECCO~1, BC4 and BC5 may belong to LVC or to Cluster A. According to \citet{boselli14} all these substructures of Virgo have the same mean distance from us. In the following we will consider all the newly confirmed BCs and SECCO~1 as members of the Virgo cluster, adopting D=16.5~Mpc \citep{mei2007} for all of them, as done in Pap-II.

\begin{figure*}[!htbp]
\gridline{\fig{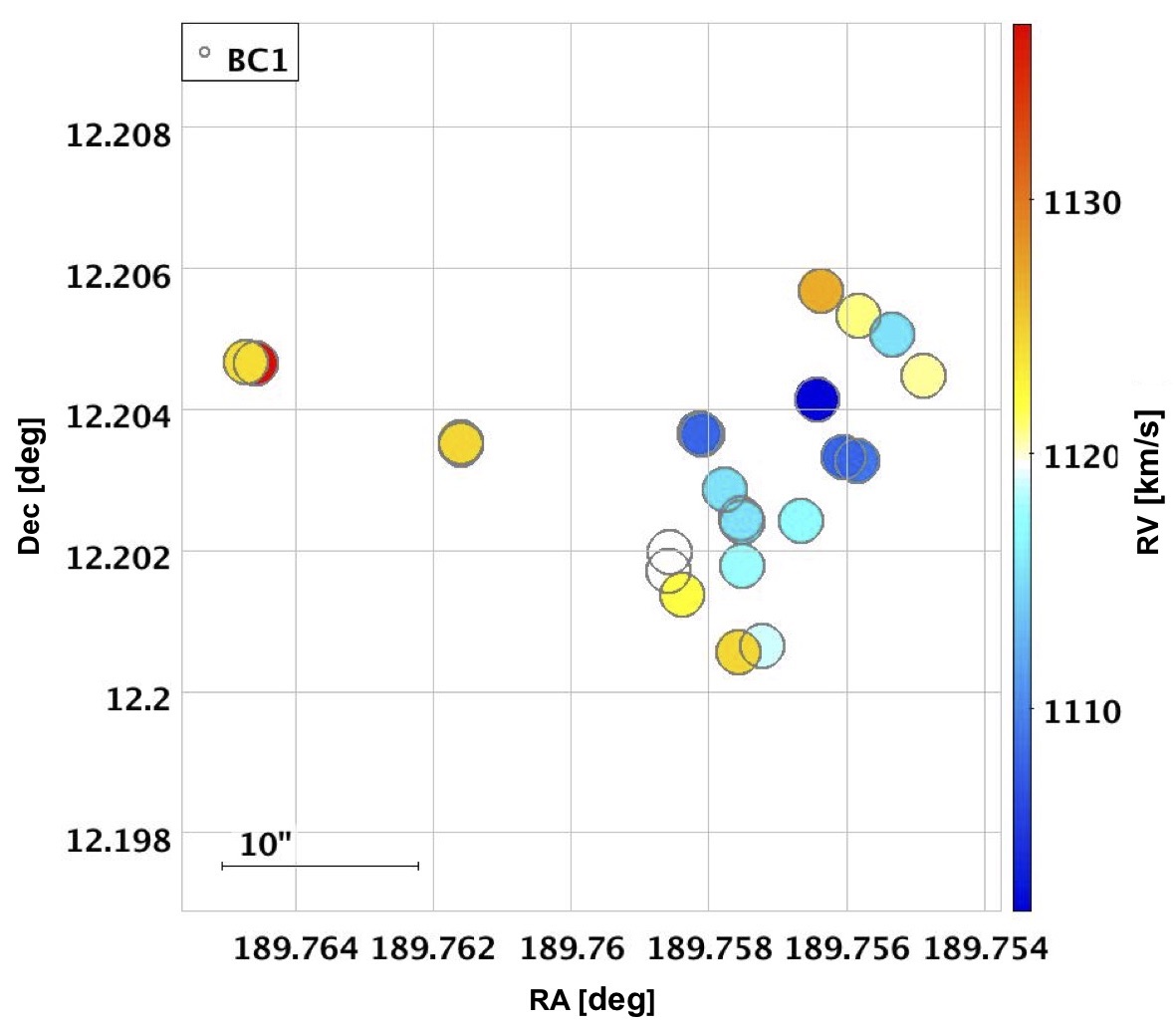}{0.41\textwidth}{(a)}
          \fig{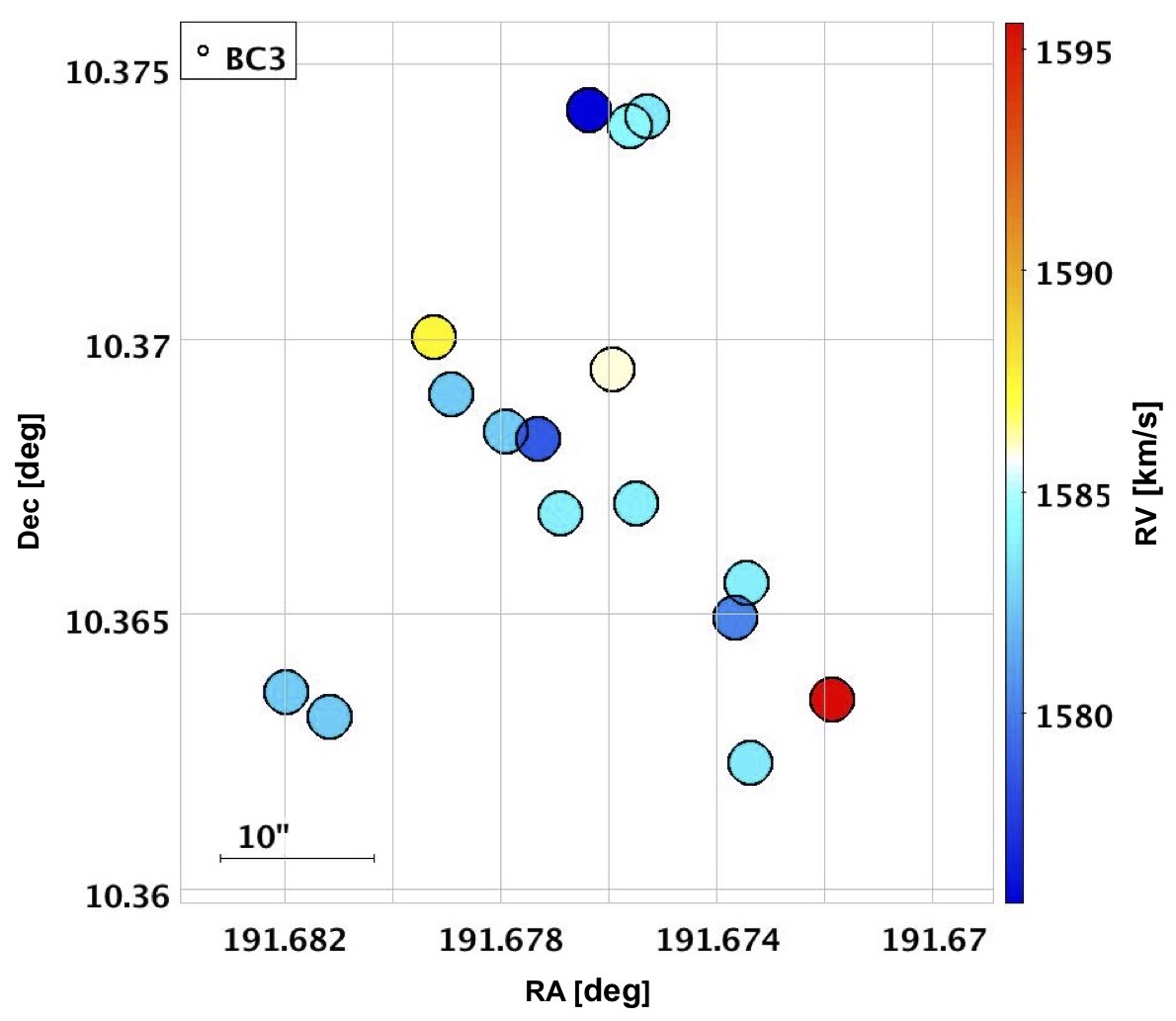}{0.41\textwidth}{(b)}
          }
\gridline{\fig{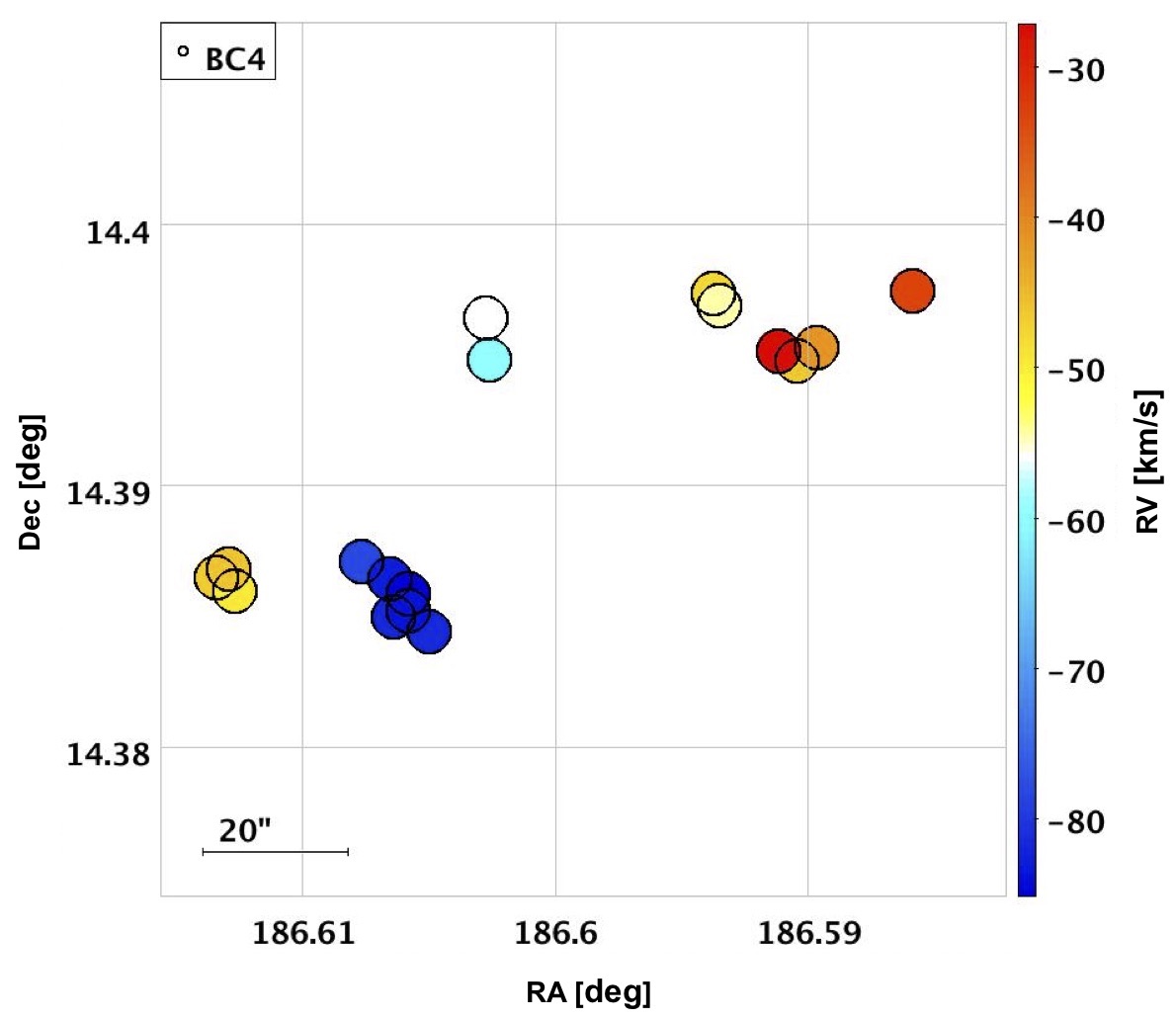}{0.41\textwidth}{(c)}
          \fig{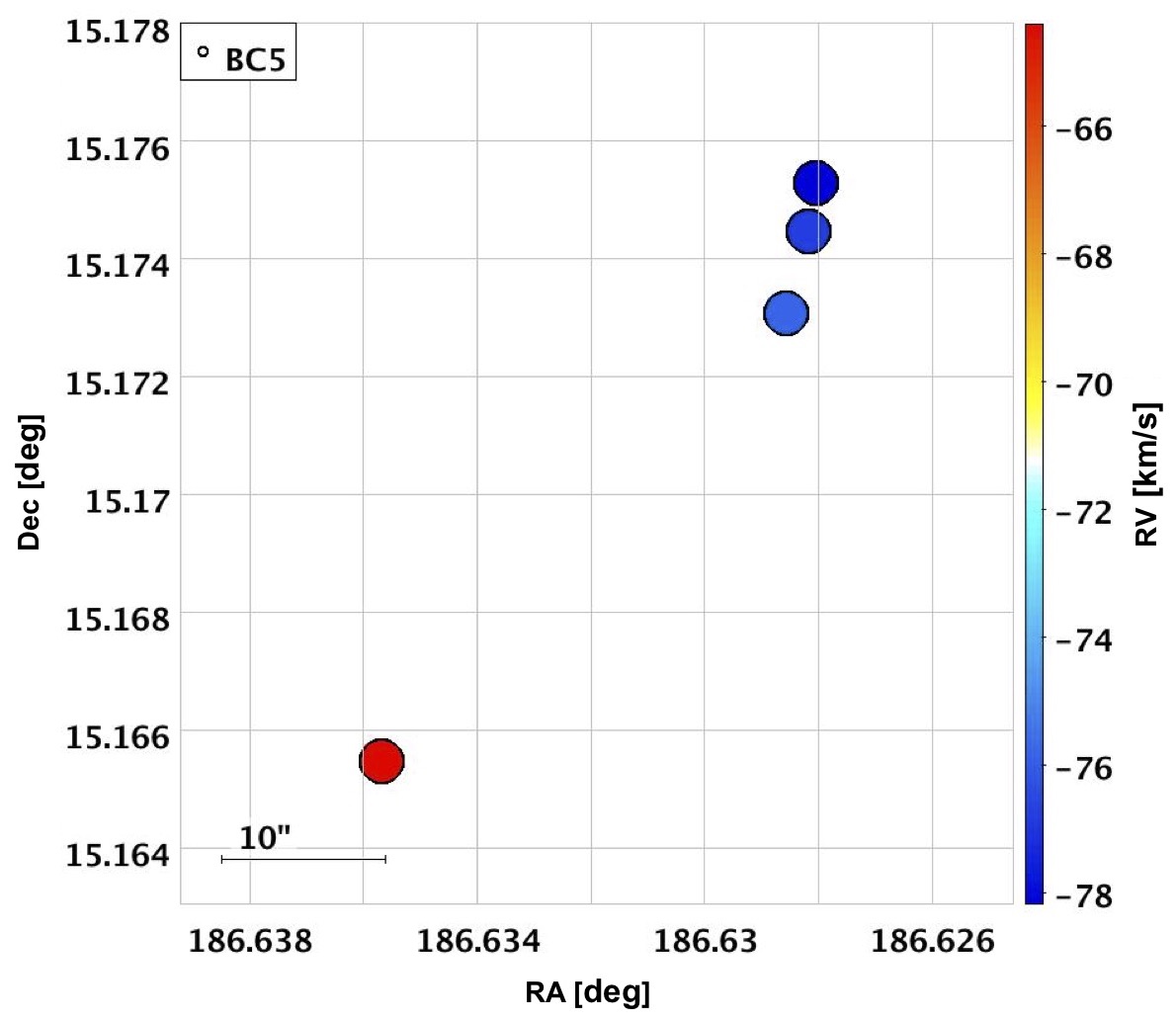}{0.41\textwidth}{(d)}
          }
\caption{Radial velocity fields, in RA [deg] and Dec [deg] of the four systems, as traced by the individual \hii sources. BC4L and BC4R sources are shown in the same map. The source at the South-Eastern corner of panel (d) is BC5s3, not shown in Fig.~\ref{fig:map45}c.
\label{fig:rvmap}}
\end{figure*}

\begin{figure}[!htbp]
\includegraphics[width=\columnwidth]{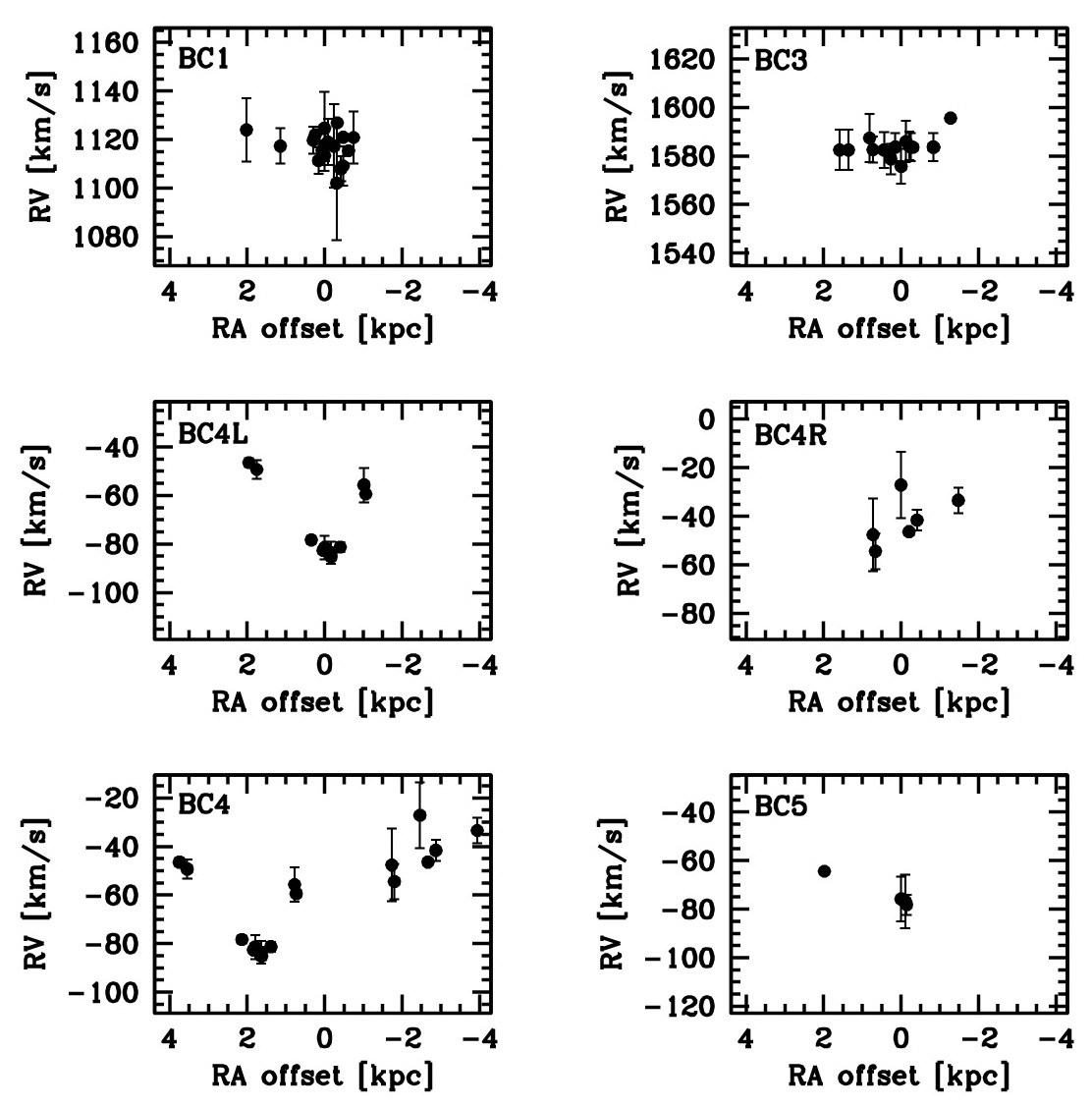}
\caption{Projected distance from the center of the system along the RA direction ({ RA offset}; East is toward the left) vs. radial velocity for BC1, BC3, BC4L, BC4R, BC4 as a whole, and BC5, from top to bottom and from left to right, respectively.}
\label{fig:rvx}
\end{figure}

In Fig.~\ref{fig:rvmap}, the maps of all the individual sources detected with {\tt Sextractor} are shown for all the considered systems, color coded according to the source RV.
BC4L and BC4R are shown in the same map; their proximity and similar RV indicate their common origin.
In each BC the star forming sources have the same RV within a few tens of $\kms$ indicating that all are part of the same system, having a common origin, which is also confirmed by their chemical homogeneity (see Sect.~\ref{sec:meta}). In all cases, some sign of kinematic coherence of sub-groups of adjacent sources is perceivable, suggesting that the systems are structured into clumps, that, possibly, are slowly flying apart one from the other (see below). To make a direct comparison between the physical size and the kinematics of the various systems, including RV uncertainties, in Fig.~\ref{fig:rvx} we plot the projected distance from the center of the system (along the right ascension direction, { RA offset}) and radial velocity of each individual source. The adopted centers are listed in Tab.~\ref{tab:means}.
BC4L and BC4R are shown separately (middle panels of Fig.~\ref{fig:rvx}) and together in a single panel (lower left panel). The observed configurations suggest different degrees of spatial and kinematic coherence, with some hints of velocity gradients. In particular the diagram showing the two pieces of BC4 together may suggest the case of a system moving toward us, lead by the dense clump around { RA offset}$=-2$~kpc, while sources to both sides of it are lagging behind proportionally to their physical distance, reminiscent of the configuration produced in the simulation with star-formation by \citet[][see their Fig.~11]{Calura20}.

\begin{figure}[!htbp]
\includegraphics[width=0.7\columnwidth]{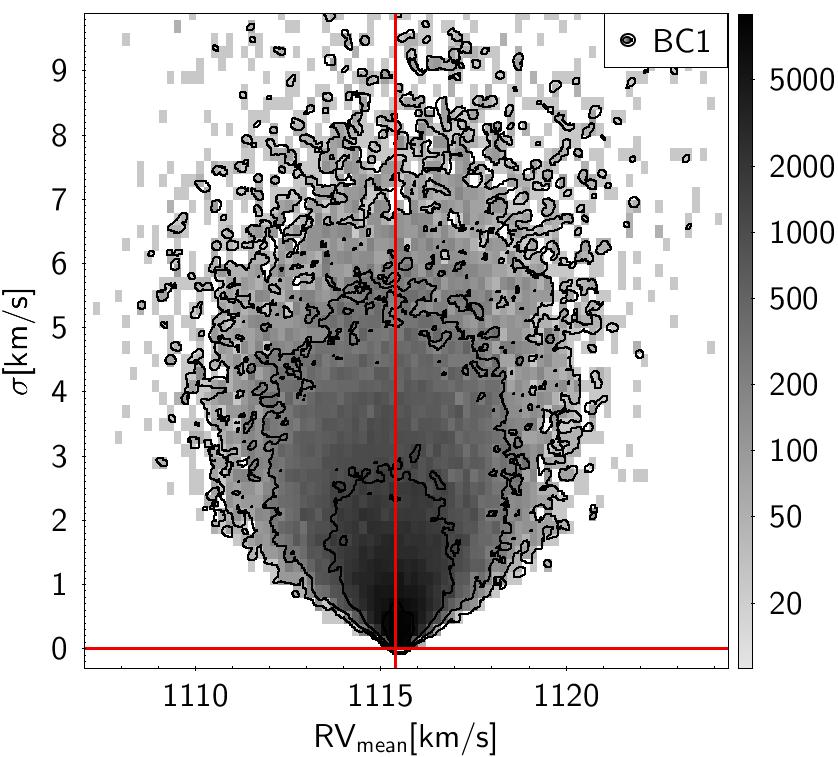}
\includegraphics[width=0.7\columnwidth]{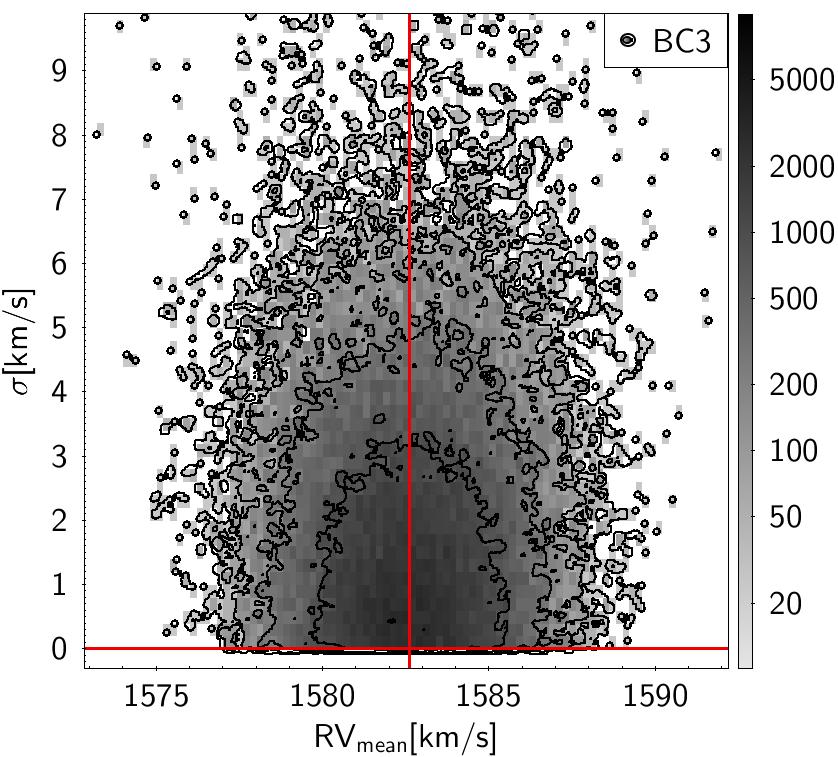}
\caption{2D posteriori PDF of $RV_{mean}$ and $\sigma_{int}$ as sampled with MCMC, for BC1 and BC3. The red vertical line marks the median of the $RV_{mean}$ distribution, while the red horizontal line marks the mode of the $\sigma_{int}$ distribution.}
\label{fig:chains}
\end{figure}

The RV distributions displayed in Fig.~\ref{fig:rvmap} and Fig.~\ref{fig:rvx} suggest that a simple velocity dispersion is probably not adequate to capture the internal kinematics of these systems. The $\sigma$ values listed in Tab.~\ref{tab:means} are standard deviations, not corrected for observational uncertainties. These are of the order of the uncertainties on individual RV estimates, suggesting that the velocity dispersions are not resolved by our data. To have a deeper insight into this problem we make an attempt to estimate the intrinsic dispersion for the two systems showing both the highest degree of kinematic coherence\footnote{It is important to recall that the kinematic coherence is observed only in RV. In principle, velocity gradients similar to those observed in BC4 may be present also in BC1 and BC3, just hidden by projection effects.} and the largest number of individual sources, BC1 and BC3. To have reliable uncertainties associated to each RV estimate we selected only sources whose RV and $\epsilon RV$ were obtained from at least three different spectral lines, thus selecting samples of 13 and 12 sources for BC1 and BC3, respectively.

From these data we derived the Probability Density Function (PDF) of the parameters of a simple Gaussian model ($V_{sys}$, $\sigma_{int}$) through a Bayesian analysis, using a Monte Carlo Markov Chain (MCMC) analysis, as done in \citet{mb_pw1}. We used {\tt JAGS}\footnote{\tt http://mcmc-jags.sourceforge.net}, within the {\tt R}\footnote{\tt https://www.r-project.org} environment, to run four independent MCMCs of 10000 steps each, after a burn in phase of 1000 steps. For both parameters uniform priors were adopted: for $V_{sys}$ in a range of $\simeq \pm 50\kms$ around the mean RV, for $\sigma_{int}$ in the range $0.0\kms< \sigma_{int} <10.0\kms$. The resulting 2D PDFs, as sampled by MCMCs, are shown in Fig.~\ref{fig:chains}. The median ($P_{50}$) $\pm$ the semi-difference between 16th and 84th percentiles of the marginalised $V_{sys}$ PDF are
$P_{50}= 1115.4 \pm 1.2\kms$ for BC1, and 
$P_{50}= 1582.6 \pm 2.1\kms$ for BC3, in agreement, within the uncertainty, with the straight averages listed in Tab.~\ref{tab:means}. On the other hand, Fig.~\ref{fig:chains} clearly demonstrates that the velocity dispersions are unresolved by our data, even in these most favourable cases. For both systems the PDF reaches its maximum at $\sigma_{int}=0.0\kms$. For BC1(BC3) half of the points sampling the PDF have $\sigma_{int}\le 1.7(1.6)\kms$, 75\% have $\sigma_{int}\le 3.0(2.8)\kms$, and 95\% have $\sigma_{int}\le 5.6(5.1)\kms$. It can be concluded that BC1 and BC3 may have $\sigma_{int}$ virtually anywhere between $\simeq 0\kms$ and $\simeq 6\kms$, but most likely $\sigma_{int}\le 3.0\kms$.

\citet{Calura20} introduced a stellar virial ratio ($\alpha_{vir}$, their Eq.~8) as a simple parameter to evaluate if a stellar system is gravitationally bound ($\alpha_{vir}\la 1$) or not ($\alpha_{vir}\gg 1$). We can use the version of their equation having the half light radius ($R_h$) as input parameter, instead of the 3D half-mass radius, to estimate $\alpha_{vir}$ for the BCs. Pap-II estimates for these systems stellar masses in the range $4\times 10^4~M_{\sun} - 1\times 10^5~M_{\sun}$. Assuming, conservatively, $M=10^5~M_{\sun}$, taking $R_{med}$ from Tab.~\ref{tab:means} as a proxy for $R_h$, and adopting $\sigma_{int}=1.0\kms$ we obtain $\alpha_{vir}=6.9, 11.7, 7.4, 6.8, 4.2$, for BC1, BC3, BC4L, BC4R, and BC5, respectively. Assuming $\sigma_{int}=2.0\kms$ will move all the $\alpha_{vir}$ to values in the range $15 - 50$, while keeping $\sigma_{int}=1.0\kms$ and assuming $M=10^6~M_{\sun}$ would imply $\alpha_{vir}\la 1$ for all the systems. We conclude that BC1, BC3, BC4L, BC4R, and BC5 are most likely unbound, as stellar systems. However they can be considered somehow borderline, given the sizeable uncertainties in all the parameters involved in the computation of $\alpha_{vir}$ and the inadequacy of a simple gaussian to model their velocity distribution. It is quite possible that some of the sub-clumps they are made of would leave a bound remnant, a small open cluster -like system floating undisturbed within Virgo, while its stars evolve passively \citep[an hypothesis already suggested by][]{secco1_18}.

\section{Metallicity and star formation}
\label{sec:meta}

Given the available lines with measured fluxes, corrected for extinction, we estimated the gas phase oxygen abundance using two different strong-line ratios, N2=[N{\sc ii}]/H$\alpha$, and O3N2=([O{\sc iii}]/H$\beta$)/([N{\sc ii}]/H$\alpha$), as defined by \citet[][PP04 hereafter]{pp04}. We were able to measure N2 for the 35 sources listed in Table~\ref{tab:met} and O3N2 for 15 of them. In Tab.~\ref{tab:met} we provide the values of $12+{\rm log}(O/H)$ derived from N2 and O3N2 using both the calibration by PP04 and by \citet[][M13 hereafter]{marino13}. 
The individual errors on the oxygen abundance of each region properly include the contribution of the uncertainties on the fluxes of the emission lines and on their correction for reddening, as well as of the uncertainty associated to the adopted calibrations. 
Following Be17, to 
compensate for the effects of varying ionisation, we compute the oxygen abundance as the average of the abundances from N2 and O3N2 (N2+03N2), taking the average from the PP04 calibrations as our preferred value. The mean abundances of the studied systems range from $12+{\rm log}(O/H)=8.29 \pm 0.10$ to $12+{\rm log}(O/H)=8.73 \pm 0.04$ (Tab.~\ref{tab:means}), clearly much larger than expected values for galaxies with such a low stellar mass, that typically have $12+{\rm log}(O/H)\le 7.5$  \citep[see. e.g.,][]{hidalgo17}. This indicates that these systems have likely originated from gas stripped from larger galaxies, like SECCO~1 (see Pap-II for a deeper discussion).

The mean abundance and standard deviations reported in Tab.~\ref{tab:means} are obtained from N2+03N2, hence they are limited to the few sources per system having estimates of O3N2. However, we can use the abundances from N2 (PP04 calibration), that are available for many more sources, to have a more realistic estimate of the uncertainties on the mean abundances and to study the chemical homogeneity of the various BCs. Unfortunately, the uncertainty on the abundance of individual sources can be quite large, ranging from $0.09$~dex to $0.9$~dex, providing poor constraints on the abundance spread, especially for BC1, where { even} the brightest sources have relatively low signal to noise spectra. As in the case of velocity dispersion, also the metallicity dispersion is not resolved by our data. However we can attempt to obtain some collective constraint on the intrinsic dispersion. 

Using the simple Maximum Likelihood algorithm described by
\cite{muccia12}, we find mean oxygen abundances of $\langle 12+{\rm log}(O/H)\rangle = 8.3\pm 0.2, 8.2\pm 0.1, 8.66\pm 0.04$, and $8.56\pm 0.07$, for BC1, BC3, BC4 and BC5, respectively. In all cases the most likely value for the intrinsic dispersion is zero, with 1$\sigma$ uncertainties ranging from $0.04$~dex (BC4, from 13 sources) to $0.2$~dex (BC1, from 10 sources).
We conclude: a) that the typical internal uncertainties for the mean abundances range from 0.04~dex to 0.2~dex, and b) that all the systems appear as remarkably homogeneous from the chemical point of view, again similar to SECCO~1. This supports the view that all the \hii regions within a given BC were born from the same gas cloud. The fact that these sources were born together and now lie in clumps between $\sim 1 - 8$ ~kpc (in projection) from each other may suggest that they are in the process of dissolving.

Finally, in Tab.~\ref{tab:means} we report also the estimates of the total integrated H$\alpha$ flux for each BC, obtained by photometry with large apertures on the continuum-subtracted images shown in Fig.~\ref{fig:map13} and Fig.~\ref{fig:map45}. Once corrected for the average extinction using the extinction law by \citet{calzetti2000} with $R_V=3.1$, the integrated fluxes can be can be converted into estimates of the current Star Formation Rate (SFR) using Eq.~2 of \cite{kennicutt98}.
SFR for the considered systems ranges from $\simeq 0.3\times 10^{-3} M_{\sun}~yr^{-1}$ (BC1, BC5) to $\simeq 1.7\times 10^{-3} M_{\sun}~yr^{-1}$ (BC3), to be compared with $\simeq 0.7\times 10^{-3} M_{\sun}~yr^{-1}$ of SECCO~1, lying in the same range \citep{secco1_18}.

\section{Summary and conclusions}
\label{sec:conclu}

We have presented the results of MUSE observations of five candidate isolated star forming regions, optically selected to be similar to the prototype of the class SECCO~1. The acquired spectra allowed us to reject one of the candidates (BC2) and to confirm the other four (BC1, BC3, BC4, and BC5) as genuine star forming regions, likely lying in the Virgo cluster of galaxies (see Pap-II for additional support to this conclusion). All the physical properties that we were able to measure are similar to those observed in SECCO~1.
In particular:

\begin{itemize}

\item in all the confirmed BCs we identified several \hii regions, plus some diffuse hot gas.

\item The mean heliocentric velocity of each BC is consistent with membership in the Virgo cluster.

\item Each BC is typically composed of a few, separated, star forming clumps, with systemic velocities within, at most, a few tens of $\kms$. In the case of the most extended system, BC4, velocity gradients suggesting ongoing disruption are observed. 

\item Our velocity estimates do not have a sufficient precision to resolve the velocity dispersion of the considered systems, that, however, should be, in all cases $\sigma \la 20~\kms$ . Still, given the available constraints, it seems unlikely that they can survive as gravitationally bound stellar systems.

\item The mean oxygen abundance of each BC is significantly larger than that expected for galaxies of similar stellar mass, strongly suggesting that they originated from gas clouds stripped from larger galaxies. Each BC appears internally homogeneous in terms of oxygen abundance, within the limits of the available observation, suggesting that all the associated sources were born from the same gas cloud.

\item The instantaneous SFR is $0.3\times 10^{-3} M_{\sun}~yr^{-1}\la {\rm SFR} \la 2.0\times 10^{-3} M_{\sun}~yr^{-1}$.

\end{itemize}

These results, together with those obtained from {\em HST} and \hi observations are discussed in the companion paper Pap-II, where an evolutionary path for the studied system is proposed.

\startlongtable
\begin{table*}
\caption{Position, radial velocity and observed  H$\alpha$ flux of individual sources}
\label{tab:RVpos}
\hskip -3cm
\begin{tabular}{lcccccccc}
\hline
  name   &    ra    &	  dec  &    RV  &   $\epsilon$RV & N$_{RV}$  &  F(H$\alpha$) &  $\epsilon$F(H$\alpha$) & FWHM \\
         &   [deg]  &	[deg]  & [km s$^{-1}$] &[km s$^{-1}$] & &[10$^{-18}$ erg cm$^{-2}$ s$^{-1}$] & [10$^{-18}$ erg cm$^{-2}$ s$^{-1}$] & arcsec\\
\hline
  BC1s11 & 189.75810 &  12.20365 & 1111.3 &  5.5 & 7 &    97.6 &    7.9 & 2.2	\\
  BC1s12 & 189.75754 &  12.20245 & 1113.4 &  6.3 & 5 &    58.8 &    5.9 & 8.2	\\
  BC1s13 & 189.76160 &  12.20350 & 1117.3 &  7.3 & 5 &    69.0 &    6.5 & 2.4	\\
  BC1s14 & 189.75489 &  12.20448 & 1120.8 & 10.8 & 4 &    23.4 &    4.2 & 1.5	\\
  BC1s15 & 189.75585 &  12.20327 & 1109.1 &  8.1 & 5 &    49.6 &    5.5 & 3.6	\\
  BC1s17 & 189.75859 &  12.20172 & 1119.7 &  5.6 & 4 &    14.1 &    3.7 & 1.2	\\
  BC1s20 & 189.75723 &  12.20065 & 1119.0 &  9.5 & 5 &    40.8 &    5.0 & 2.9	\\
 BC1s57W$^{\dagger}$ & 189.75637 &  12.20567 & 1126.9 & 20.0 & 1 &  9.4 & 3.0 & 1.3	\\ 
 BC1s61W$^{\dagger}$ & 189.75583 &  12.20534 & 1120.9 & 20.0 & 1 &  8.6 & 3.0 & 0.5	\\ 
 BC1s62W$^{\dagger}$ & 189.75534 &  12.20506 & 1115.4 & 20.0 & 1 &     8.4 &    3.4 & 0.6	\\ 
 BC1s65W$^{\dagger}$ & 189.76471 &  12.20468 & 1124.0 & 13.0 & 3 &    11.7 &    3.6 & 1.7	\\ 
 BC1s67W$^{\dagger}$ & 189.75644 &  12.20414 & 1102.0 & 23.4 & 2 &     8.3 &    3.4 & 1.5	\\ 
 BC1s73W & 189.75604 &  12.20332 & 1108.0 &  5.2 & 5 &    43.6 &    5.2 & 2.6	\\
 BC1s80W & 189.75777 &  12.20287 & 1115.5 &  2.0 & 4 &    57.4 &    5.9 & 4.0	\\
 BC1s83W & 189.75667 &  12.20241 & 1117.3 & 17.2 & 4 &    13.5 &    3.7 & 0.9	\\
 BC1s90W & 189.75751 &  12.20178 & 1117.7 &  6.6 & 4 &    55.0 &    5.8 & 1.7	\\
 BC1s92W & 189.75840 &  12.20138 & 1121.9 & 20.0 & 1 &    15.3 &    3.8 & 0.5	\\
 BC1s98W & 189.75757 &  12.20055 & 1124.6 & 15.1 & 4 &    29.5 &    4.5 & 1.4	\\
   BC3s4 & 191.67637 &  10.37416 & 1575.7 &  7.2 & 5 &   219.7 &   14.0 & 5.3	\\
   BC3s9 & 191.67924 &  10.37005 & 1587.4 &  9.9 & 4 &   167.2 &   11.4 & 2.7	\\
  BC3s10 & 191.67892 &  10.36901 & 1582.6 &  5.3 & 7 &   465.3 &   26.3 & 1.9	\\
  BC3s12 & 191.67594 &  10.36945 & 1586.0 &  8.5 & 4 &    33.4 &    4.7 & 2.4	\\
  BC3s13 & 191.67792 &  10.36832 & 1582.5 &  7.4 & 5 &    62.6 &    6.1 & 1.7	\\
  BC3s14 & 191.67730 &  10.36820 & 1578.7 &  6.2 & 3 &    24.9 &    4.2 & 1.5	\\
  BC3s15 & 191.67549 &  10.36701 & 1583.7 &  5.8 & 7 &   390.4 &   22.5 & 3.6	\\
  BC3s16 & 191.67691 &  10.36682 & 1583.7 &  5.8 & 7 &    63.1 &    6.2 & 4.3	\\
  BC3s18 & 191.67344 &  10.36558 & 1583.7 &  5.8 & 7 &   489.6 &   27.5 & 1.6	\\
  BC3s19 & 191.68198 &  10.36358 & 1582.5 &  8.3 & 5 &    68.4 &    6.4 & 5.2	\\
  BC3s20 & 191.68117 &  10.36314 & 1582.5 &  8.3 & 5 &    53.0 &    5.7 & 3.3	\\
  BC3s23 & 191.67339 &  10.36230 & 1583.6 & 20.0 & 1 &    12.5 &    3.6 & 0.7	\\
 BC3s24W & 191.67529 &  10.37407 & 1583.6 & 20.0 & 1 &    27.4 &    4.4 & 1.4	\\
 BC3s26W & 191.67561 &  10.37388 & 1584.1 &  5.8 & 4 &    49.6 &    5.5 & 1.2	\\
 BC3s69W & 191.67186 &  10.36344 & 1595.6 & 20.0 & 1 &    12.5 &    3.5 & 1.1	\\
  BC4s3L & 186.60278 &  14.39644 &  -55.7 &  7.1 & 5 &    10.7 &    3.5 & 1.1	\\
  BC4s4L & 186.60263 &  14.39481 &  -59.4 &  1.1 & 5 &   121.9 &    9.1 & 1.2	\\
  BC4s8L & 186.60766 &  14.38708 &  -78.3 &  1.7 & 7 &   143.3 &   10.2 & 1.2	\\
  BC4s9L & 186.60658 &  14.38645 &  -82.6 &  1.3 & 6 &    89.0 &    7.4 & 1.8	\\
 BC4s10L & 186.60584 &  14.38589 &  -85.2 &  1.6 & 6 &   149.8 &   10.5 & 2.6	\\
 BC4s11L & 186.61344 &  14.38647 &  -46.5 &  1.8 & 6 &    49.8 &    5.5 & 1.5	\\
 BC4s12L & 186.61272 &  14.38597 &  -49.3 &  3.9 & 6 &    44.9 &    5.2 & 4.2	\\
 BC4s13L & 186.60643 &  14.38496 &  -81.4 &  4.9 & 3 &     6.7 &    3.3 & 1.2	\\
 BC4s14L & 186.60498 &  14.38439 &  -81.3 &  2.0 & 3 &    17.8 &    3.9 & 2.1	\\
 continue   &           &   &   &   &  &     &     & 	\\
\hline
\hline
\end{tabular}
\end{table*}


\begin{table*}
\tablenum{2}
\caption{continued}
\label{tab:RVpos}
\hskip -3cm
\begin{tabular}{lcccccccc}
\hline
  name   &    ra    &	  dec  &    RV  &   $\epsilon$RV & N$_{RV}$  &  F(H$\alpha$) &  $\epsilon$F(H$\alpha$) & FWHM \\
         &   [deg]  &	[deg]  & [km s$^{-1}$] &[km s$^{-1}$] & &[10$^{-18}$ erg cm$^{-2}$ s$^{-1}$] & [10$^{-18}$ erg cm$^{-2}$ s$^{-1}$] & arcsec\\
\hline
BC4s38WL & 186.60586 &  14.38522 &  -83.6 &  4.6 & 5 &    19.7 &    4.0 & 1.6	\\
 BC4s11R & 186.58587 &  14.39747 &  -33.4 &  5.2 & 5 &     8.1 &    3.4 & 0.9	\\
 BC4s12R & 186.59377 &  14.39739 &  -47.6 & 15.0 & 3 &    11.4 &    3.6 & 0.9	\\
 BC4s15R & 186.59044 &  14.39478 &  -46.4 &  1.3 & 6 &    67.7 &    6.4 & 2.8	\\
BC4s34WR & 186.59354 &  14.39690 &  -54.4 &  7.3 & 2 &     4.8 &    3.2 & 1.5	\\
BC4s40WR & 186.58970 &  14.39530 &  -41.6 &  4.3 & 5 &     8.9 &    3.4 & 0.8	\\
BC4s41WR & 186.59118 &  14.39515 &  -27.1 & 13.6 & 2 &     8.4 &    3.4 & 1.3	\\
BC5s3$^{\star}$ & 186.63569 &  15.16547 &  -64.4 & 20.0 & 1 &    39.7 &    5.0 &.1.1	\\ 
BC5s9 & 186.62805 &  15.17529 &  -78.2 &  4.1 & 5 &   111.7 &    8.6 & 2.9	\\
BC5s10 & 186.62818 &  15.17447 &  -76.8 & 11.1 & 5 &    73.0 &    6.6 & 1.4	\\
BC5s12 & 186.62856 &  15.17307 &  -75.8 &  9.2 & 6 &   213.0 &   13.7 & 4.1	\\
\hline
\hline
\end{tabular}
\tablecomments{{ Columns description: name of the source, right ascension (J2000), declination (J2000), heliocentric radial velocity and its uncertainty ($\epsilon$RV), H$\alpha$ flux and its uncertainty ($\epsilon$F(H$\alpha$)), and the Full Width at Half Maximum as measured by {\tt Sextractor}.}
$^{\dagger}$ These sources may be particularly affected by contamination from the diffuse emission associated to BC1. $^{\star}$ Source located at the edge of the field of view, only partially included in the data-cube. The missing numbers in the nomenclature of the individual sources (like, e.g, BC1s1 to BC1s10, or BC3s5 to BC3s8), correspond to sources that were detected by {\tt Sextractor} but did not passed the selection by visual inspection of the spectra described in Sect.~\ref{sec:obse}. F(H$\alpha$) is the observed flux, not corrected for extinction (see Tab.~\ref{tab:lineFlux}).}
\end{table*}

\begin{table*}
\caption{Line fluxes of individual sources, in units of the H$\beta$ flux, set to F(H$\beta$)=100.}
\label{tab:lineFlux}
\hskip -3cm
\begin{tabular}{lccccccc}
\hline
\hline
  Name     & F(H$\beta$)    & [OIII]$_{5007}$      & H$\alpha$	  &  [NII]$_{6584}$ & [SII]$_{6717}$	& [SII]$_{6731}$     & C$_{\beta}$     \\
           &[10$^{-18}$ erg cm$^{-2}$ s$^{-1}$] &  &              &		    &  		        &                    &		mag       \\
  \hline         
  BC1s11   &  8.1 $\pm$ 3.4 & 138.3 $\pm$45.2	& 299.4$\pm$97.2  &  24.3$\pm$ 42.0 &  49.8$\pm$ 48.0	&  34.7$\pm$	44.7 &1.99$\pm$0.21    \\
  BC1s12   & 13.6 $\pm$ 3.7 &      \nodata      & 290.5$\pm$43.7  &  23.1$\pm$ 23.8 &  73.1$\pm$ 27.7	&  49.6$\pm$	25.9 &0.57$\pm$0.16    \\
  BC1s13   & 21.8 $\pm$ 4.1 &      \nodata      & 287.8$\pm$29.6  &  19.0$\pm$ 14.8 &  94.1$\pm$ 19.0	&  61.6$\pm$	17.2 &0.14$\pm$0.12    \\
  BC1s14   &  6.9 $\pm$ 3.3 &      \nodata      & 288.4$\pm$60.3  &  37.5$\pm$ 45.5 &  91.0$\pm$ 48.7	&  51.8$\pm$	46.4 &0.23$\pm$0.28    \\
  BC1s15   & 11.9 $\pm$ 3.6 &      \nodata      & 290.2$\pm$46.0  &  17.2$\pm$ 26.4 & 103.9$\pm$ 32.8	&  76.5$\pm$	30.8 &0.52$\pm$0.18    \\
  BC1s17   &  4.0 $\pm$ 3.2 &      \nodata      & 288.8$\pm$92.8  &      \nodata      &      \nodata    &      \nodata       &0.29$\pm$0.46    \\
  BC1s20   & 13.1 $\pm$ 3.7 &      \nodata      & 287.7$\pm$38.4  &  19.0$\pm$ 23.8 &  66.1$\pm$ 26.4	&  45.7$\pm$	25.3 &0.11$\pm$0.17    \\
  BC1s62W  &  6.4 $\pm$ 3.3 &      \nodata      & 131.1$\pm$53.5  &      \nodata      &      \nodata    &      \nodata       &0.00$\pm$0.40    \\
  BC1s73W  &  7.3 $\pm$ 3.4 &      \nodata      & 293.3$\pm$70.8  &  18.4$\pm$ 42.9 & 109.6$\pm$ 52.7	&  68.6$\pm$	48.4 &1.01$\pm$0.25    \\
  BC1s80W  & 12.6 $\pm$ 3.6 & 39.4  $\pm$25.9	& 291.0$\pm$46.6  &  22.1$\pm$ 25.6 &  65.4$\pm$ 29.1	&  42.6$\pm$	27.3 &0.64$\pm$0.17    \\
  BC1s83W  &  3.7 $\pm$ 3.2 &      \nodata      & 288.9$\pm$98.0  &  10.0$\pm$ 80.7 & 170.1$\pm$ 90.8	&  93.3$\pm$	86.0 &0.31$\pm$0.48    \\
  BC1s90W  & 12.3 $\pm$ 3.6 &      \nodata      & 290.8$\pm$46.7  &  12.3$\pm$ 25.3 &  80.5$\pm$ 30.7	&  58.1$\pm$	29.0 &0.61$\pm$0.17    \\
  BC1s92W  &  5.0 $\pm$ 3.2 &      \nodata      & 287.6$\pm$75.3  &  33.7$\pm$ 61.7 &  42.9$\pm$ 62.2	&  34.2$\pm$	61.8 &0.09$\pm$0.38    \\
  BC1s98W  & 12.3 $\pm$ 3.6 &      \nodata      & 239.8$\pm$36.4  &  12.3$\pm$ 25.0 &  66.0$\pm$ 27.7	&  40.1$\pm$	26.4 &0.00$\pm$0.19    \\
  BC3s4    & 70.3 $\pm$ 6.5 &      \nodata      & 287.7$\pm$19.9  &  10.8$\pm$  4.8 &  52.1$\pm$  7.1	&  34.8$\pm$	 6.2 &0.12$\pm$0.07    \\
  BC3s9    & 49.6 $\pm$ 5.5 & 14.7  $\pm$6.8	& 288.4$\pm$22.9  &  12.3$\pm$  6.8 &  33.2$\pm$  8.0	&  19.6$\pm$	 7.2 &0.22$\pm$0.08    \\
  BC3s10   &143.1 $\pm$10.1 & 37.1  $\pm$4.0	& 288.1$\pm$18.3  &  12.8$\pm$  2.8 &  27.4$\pm$  3.7	&  20.3$\pm$	 3.2 &0.17$\pm$0.05    \\
  BC3s12   &  6.9 $\pm$ 3.3 &      \nodata      & 291.5$\pm$67.7  &  \nodata &  \nodata	&  \nodata &0.73$\pm$0.27    \\
  BC3s13   & 13.4 $\pm$ 3.7 &      \nodata      & 291.1$\pm$45.6  &  16.6$\pm$ 23.6 &  32.4$\pm$ 25.0	&  22.7$\pm$	24.2 &0.67$\pm$0.16    \\
  BC3s15   &124.8 $\pm$ 9.2 & 195.1 $\pm$12.3	& 287.7$\pm$18.0  &   9.5$\pm$  2.9 &  23.0$\pm$  3.7	&  19.2$\pm$	 3.4 &0.12$\pm$0.06    \\
  BC3s16   & 13.8 $\pm$ 3.7 & 48.4  $\pm$24.2	& 291.0$\pm$44.5  &  23.2$\pm$ 23.5 &  51.7$\pm$ 25.9	&  36.7$\pm$	24.7 &0.64$\pm$0.16    \\
  BC3s18   & 47.5 $\pm$ 5.4 &      \nodata      & 298.1$\pm$57.9  &  10.5$\pm$  8.1 &  24.4$\pm$ 10.9	&  16.7$\pm$	 9.5 &1.77$\pm$0.07    \\
  BC3s19   & 24.9 $\pm$ 4.2 &      \nodata      & 274.8$\pm$25.8  &  15.8$\pm$ 12.8 &  34.2$\pm$ 13.8	&  24.7$\pm$	13.3 &0.00$\pm$0.11    \\
  BC3s20   &  8.7 $\pm$ 3.4 &      \nodata      & 293.4$\pm$64.7  &   1.8$\pm$ 34.5 &  44.8$\pm$ 39.2	&  25.3$\pm$	37.1 &1.04$\pm$0.21    \\
  BC3s26W  & 14.4 $\pm$ 3.7 & 44.1  $\pm$23.0	& 288.5$\pm$37.9  &  15.9$\pm$ 21.7 &  47.7$\pm$ 23.6	&  51.7$\pm$	23.9 &0.25$\pm$0.16    \\
  BC4s4L   & 31.0 $\pm$ 4.6 & 15.4  $\pm$10.5	& 289.7$\pm$29.3  & 108.4$\pm$ 17.0 &  31.9$\pm$ 11.9	&  21.9$\pm$	11.2 &0.43$\pm$0.09    \\
  BC4s8L   & 38.6 $\pm$ 4.9 & 46.6  $\pm$10.2	& 289.3$\pm$26.3  & 109.5$\pm$ 14.8 &  33.5$\pm$ 10.0	&  24.4$\pm$	 9.4 &0.36$\pm$0.08    \\
  BC4s9L   & 20.9 $\pm$ 4.0 & 23.5  $\pm$15.6	& 290.5$\pm$35.6  & 104.0$\pm$ 22.0 &  39.8$\pm$ 17.3	&  28.2$\pm$	16.5 &0.54$\pm$0.12    \\
  BC4s10L  & 39.0 $\pm$ 4.9 & 5.9   $\pm$ 8.0	& 289.6$\pm$26.9  & 103.5$\pm$ 14.6 &  27.6$\pm$  9.6	&  33.3$\pm$	10.0 &0.40$\pm$0.08    \\
  BC4s11L  &  9.8 $\pm$ 3.5 & 111.7 $\pm$36.5	& 292.0$\pm$55.9  & 152.1$\pm$ 43.8 &  53.4$\pm$ 35.4	&  33.0$\pm$	33.5 &0.79$\pm$0.20    \\
  BC4s12L  & 12.0 $\pm$ 3.6 & 41.6  $\pm$27.1	& 289.3$\pm$43.7  & 142.6$\pm$ 34.2 &  81.3$\pm$ 30.3	&  51.4$\pm$	28.4 &0.37$\pm$0.18    \\
  BC4s14L  &  5.7 $\pm$ 3.3 &      \nodata      & 287.8$\pm$68.8  &  44.6$\pm$ 55.5 &  28.6$\pm$ 54.6	&  19.8$\pm$	54.1 &0.13$\pm$0.34    \\
  BC4s38WL &  4.2 $\pm$ 3.2 &      \nodata      & 291.3$\pm$94.2  &  91.9$\pm$ 78.3 &  60.3$\pm$ 75.7	&  35.2$\pm$	73.8 &0.67$\pm$0.41    \\
  BC4s15R  & 17.3 $\pm$ 3.9 &      \nodata      & 289.7$\pm$36.9  &  85.6$\pm$ 23.1 &  38.1$\pm$ 19.9	&  19.1$\pm$	35.0 &0.43$\pm$0.14    \\
  BC5s9    & 29.4 $\pm$ 4.5 & 16.4  $\pm$11.0	& 289.4$\pm$29.2  &  69.9$\pm$ 14.8 &  39.3$\pm$ 12.8	&  26.3$\pm$	12.0 &0.39$\pm$0.10    \\
  BC5s10   & 18.2 $\pm$ 3.9 &      \nodata      & 289.9$\pm$36.6  &  23.7$\pm$ 18.1 &  91.4$\pm$ 23.0	&  60.5$\pm$	20.79&0.46$\pm$0.13    \\
  BC5s12   & 56.7 $\pm$ 5.8 & 9.4   $\pm$5.8	& 289.3$\pm$24.1  &  77.7$\pm$ 10.3 &  46.2$\pm$  8.3	&  29.5$\pm$	7.24 &0.37$\pm$0.07    \\
\hline
\hline
\end{tabular}
\tablecomments{Notes: F(H$\beta$) is the observed flux while all other (normalised) line fluxes are corrected for extinction $C_{\beta}$.}
\end{table*}


\begin{table*}
\caption{Metallicity of individual sources.}
\label{tab:met}
\hskip -3cm
\begin{tabular}{lcccccc}
\hline
\hline
  Name      &     12+{\rm log}(O/H)   & 12+{\rm log}(O/H)	 & $\langle{\rm 12+{\rm log}(O/H)}\rangle$ &12+{\rm log}(O/H)   &12+{\rm log}(O/H)   &  $\langle{\rm 12+{\rm log}(O/H)}\rangle$\\    		
            &     N2(PP04)      & O3N2(PP04)	 & N2+O3N2(PP04)                     & N2(M13)      &  O3N2(M13)   &  N2+O3N2(M13)		      \\
 \hline 
  BC1s11    &	  8.28$\pm$0.88 & 8.33$\pm$ 1.02 &  8.31$\pm$ 0.90 & 8.24$\pm$ 0.88& 8.27$\pm$ 1.02& 8.26$\pm$ 0.89 \\
  BC1s12    &	  8.27$\pm$0.50 & \nodata &    \nodata & 8.24$\pm$ 0.50& \nodata  &	\nodata \\	
  BC1s13    &	  8.23$\pm$0.37 & \nodata &    \nodata & 8.20$\pm$ 0.37& \nodata  &	\nodata \\	
  BC1s14    &	  8.39$\pm$0.61 & \nodata &    \nodata & 8.33$\pm$ 0.61& \nodata  &	\nodata \\	
  BC1s15    &	  8.20$\pm$0.72 & \nodata &    \nodata & 8.18$\pm$ 0.72& \nodata  &	\nodata \\	
  BC1s20    &	  8.23$\pm$0.59 & \nodata &    \nodata & 8.20$\pm$ 0.59& \nodata  &	\nodata \\	
  BC1s80W   &	  8.26$\pm$0.56 & 8.50$\pm$ 0.84 &  8.38$\pm$ 0.60 & 8.23$\pm$ 0.56& 8.38$\pm$ 0.84& 8.30$\pm$ 0.59 \\
  BC1s90W   &	  8.12$\pm$0.95 & \nodata &    \nodata & 8.11$\pm$ 0.95& \nodata  &	\nodata \\	
  BC1s92W   &	  8.37$\pm$0.90 & \nodata &    \nodata & 8.31$\pm$ 0.90& \nodata  &	\nodata \\	
  BC1s98W   &	  8.17$\pm$0.93 & \nodata &    \nodata & 8.15$\pm$ 0.93& \nodata  &	\nodata \\	
  BC3s4     &	  8.09$\pm$0.22 & \nodata &    \nodata & 8.09$\pm$ 0.22& \nodata  &	\nodata \\	
  BC3s9     &	  8.12$\pm$0.27 & 8.55$\pm$ 0.47 &  8.34$\pm$ 0.33 & 8.11$\pm$ 0.27& 8.42$\pm$ 0.47& 8.26$\pm$ 0.32 \\
  BC3s10    &	  8.13$\pm$0.12 & 8.43$\pm$ 0.16 &  8.28$\pm$ 0.23 & 8.12$\pm$ 0.12& 8.34$\pm$ 0.16& 8.23$\pm$ 0.20 \\
  BC3s13    &	  8.19$\pm$0.67 & \nodata &    \nodata & 8.17$\pm$ 0.67& \nodata  &	\nodata \\	
  BC3s15    &	  8.06$\pm$0.15 & 8.16$\pm$ 0.18 &  8.11$\pm$ 0.25 & 8.06$\pm$ 0.15& 8.16$\pm$ 0.18& 8.11$\pm$ 0.23 \\
  BC3s16    &	  8.28$\pm$0.50 & 8.48$\pm$ 0.71 &  8.38$\pm$ 0.54 & 8.24$\pm$ 0.50& 8.37$\pm$ 0.71& 8.30$\pm$ 0.52 \\
  BC3s18    &	  8.07$\pm$0.41 & \nodata &    \nodata & 8.07$\pm$ 0.41& \nodata  &	\nodata \\	
  BC3s19    &	  8.19$\pm$0.38 & \nodata &    \nodata & 8.17$\pm$ 0.38& \nodata  &	\nodata \\	
  BC3s26W   &	  8.18$\pm$0.64 & 8.44$\pm$ 0.86 &  8.31$\pm$ 0.67 & 8.16$\pm$ 0.64& 8.34$\pm$ 0.86& 8.25$\pm$ 0.66 \\
  BC4s4L    &	  8.65$\pm$0.11 & 8.85$\pm$ 0.40 &  8.75$\pm$ 0.22 & 8.54$\pm$ 0.11& 8.61$\pm$ 0.4 & 8.58$\pm$ 0.20 \\  
  BC4s8L    &	  8.66$\pm$0.09 & 8.70$\pm$ 0.19 &  8.68$\pm$ 0.22 & 8.54$\pm$ 0.09& 8.51$\pm$ 0.19& 8.53$\pm$ 0.19 \\  
  BC4s9L    &	  8.64$\pm$0.14 & 8.78$\pm$ 0.42 &  8.71$\pm$ 0.24 & 8.53$\pm$ 0.14& 8.57$\pm$ 0.42& 8.55$\pm$ 0.22 \\  
  BC4s10L   &	  8.64$\pm$0.1  & 8.97$\pm$ 0.68 &  8.81$\pm$ 0.22 & 8.53$\pm$ 0.1 & 8.69$\pm$ 0.68& 8.61$\pm$ 0.19 \\  
  BC4s11L   &	  8.74$\pm$0.20 & 8.62$\pm$ 0.34 &  8.68$\pm$ 0.28 & 8.61$\pm$ 0.20& 8.46$\pm$ 0.34& 8.53$\pm$ 0.26 \\  
  BC4s12L   &	  8.72$\pm$0.16 & 8.75$\pm$ 0.44 &  8.73$\pm$ 0.26 & 8.60$\pm$ 0.16& 8.54$\pm$ 0.44& 8.57$\pm$ 0.23 \\  
  BC4s13L   &	  8.79$\pm$0.12 & \nodata &    \nodata & 8.66$\pm$ 0.12& \nodata  & \nodata   \\  
  BC4s14L   &	  8.44$\pm$0.63 & \nodata &    \nodata & 8.37$\pm$ 0.63& \nodata  & \nodata   \\  
  BC4s38WL  &	  8.61$\pm$0.12 & \nodata &    \nodata & 8.51$\pm$ 0.12& \nodata  & \nodata   \\
  BC4s3L    &	  8.60$\pm$0.12 & \nodata &    \nodata & 8.50$\pm$ 0.12& \nodata  & \nodata   \\
  BC4s15R   &	  8.60$\pm$0.17 & \nodata &    \nodata & 8.50$\pm$ 0.17& \nodata  & \nodata   \\
  BC4s34WR  &	  8.59$\pm$0.12 & \nodata &    \nodata & 8.49$\pm$ 0.12& \nodata  & \nodata   \\
  BC4s40WR  &	  8.67$\pm$0.12 & \nodata &    \nodata & 8.56$\pm$ 0.12& \nodata  & \nodata   \\
  BC5s9     &	  8.55$\pm$0.13 & 8.78$\pm$ 0.42 &  8.66$\pm$ 0.24 & 8.46$\pm$ 0.13& 8.56$\pm$ 0.42& 8.51$\pm$ 0.21 \\
  BC5s10    &	  8.28$\pm$0.38 & \nodata &    \nodata & 8.24$\pm$ 0.38& \nodata  & \nodata   \\  
  BC5s12    &	  8.57$\pm$0.09 & 8.87$\pm$ 0.35 &  8.72$\pm$ 0.22 & 8.48$\pm$ 0.09& 8.62$\pm$ 0.35& 8.55$\pm$ 0.19 \\
\hline
\hline
\end{tabular}
\tablecomments{Only abundance estimates with uncertainties $<1.0$~dex have been retained in this table. Individual uncertainties include also the uncertainty associated to the calibrating relations.
N$_{O/H}$: number of sources used to compute the average oxygen abundance.}
\end{table*}


\begin{table*}
\caption{Mean properties of the studied targets}
\label{tab:means}
\hskip -3cm
\begin{tabular}{lccccccccccc}
\hline
\hline
name &RA$_{J2000}$&Dec$_{J2000}$&R$_{med}$&R$_{max}$&F$_{int}$(H$\alpha$)&RV&$\sigma_{RV}$&N$_{RV}$&$\langle12+{\rm log}(O/H)\rangle$& $\sigma_{(O/H)}$& N$_{(O/H)}$\\
     &   deg      & deg         &arcsec   &arcsec &erg cm$^{-2}$ s$^{-1}$&   km s$^{-1}$& km s$^{-1}$ &  &  & &\\
\hline 
 BC1 & 189.75754& 12.20332 &   8.3 & 28.1  &	 5.5E-16 &	 1117 & 6 & 18  &8.35 &  0.04(0.2) & 2(10) \\ 
 BC3 & 191.67637& 10.36820 &  14.0 & 25.9  &	31.3E-16 &	 1584 & 4 & 15  &8.29 &  0.09(0.1) & 5(9) \\ 
BC4L & 186.60643& 14.38645 &  8.9 &  38.1 &     18.9E-16 &        -70 &16 & 10  &8.73 &  0.04(0.05) & 6(10) \\ 
BC4R & 186.59117& 14.39690 &  8.2 &  18.6 &      4.0E-16 &        -42 &10 &  6  & \nodata    &   (0.08)    & 0(3) \\
 BC4 & 186.60000& 14.39000 &  33.5 & 56.1  &	23.0E-16 &	  -60 &20 & 16  &8.73 &  0.04(0.04) & 6(13) \\ 
 BC5 & 186.62856& 15.17447 &   5.0 & 40.8  &	 5.7E-16 &	  -74 & 6 &  4  &8.70 &  0.03(0.08) & 2(3) \\ 
\hline
\hline
\end{tabular}
\tablecomments{Cooordinates: median of RA and Dec of the individual sources, 
except for BC4, where the position of the center has been estimated by eye. 
R$_{med}$: median angular distance from the center R; R$_{max}$: angular distance from the center of the outermost source.
Flux$_{int}$(H$\alpha$): total integrated observed H$\alpha$ flux, from aperture
photometry on continuum-subtracted H$\alpha$ slices of the cubes.
RV and $\sigma_{RV}$: mean heliocentric radial velocity and velocity
dispersion; N$_{RV}$ number of sources used to get median positions and mean
velocities. $\langle12+{\rm log}(O/H)\rangle$ and $\sigma_{O/H}$: average oxygen abundance and standard deviation.  The adopted individual abundance values are the mean from the N2 and O3N2 indicators according to the PP04 calibration. N$_{O/H}$: number of sources used to compute the average oxygen abundance. The numbers reported in parentheses in the $\sigma_{O/H}$ and N$_{O/H}$ are the intrinsic dispersion, as computed with the ML algorithm (Sect.~\ref{sec:meta} and the number of sources involved in the estimate.}
\end{table*}


\begin{acknowledgments}
We are grateful to an anonymous Referee for useful comments and suggestions that improved the clarity of the paper.
Based on observations collected at the European Southern Observatory under ESO programme 0101.B-0376A.
MB acknowledges the financial support of INAF - OAS Bologna. 
DJS acknowledges support from NSF grants AST-1821967 and 1813708.
AK acknowledges financial support from the State Agency for Research of the Spanish Ministry of Science, Innovation and Universities through the "Center of Excellence Severo Ochoa" awarded to the Instituto de Astrof\'{i}sica de Andaluc\'{i}a (SEV-2017-0709) and through the grant POSTDOC\_21\_00845 financed from the budgetary program 54a Scientific Research and Innovation of the Economic Transformation, Industry, Knowledge and Universities Council of the Regional Government of Andalusia.
EAKA is supported by the WISE research programme, which is financed by the Dutch Research Council (NWO). 
R. R. M. gratefully acknowledges support by the ANID BASAL project FB210003
KS acknowledges support from the Natural Sciences and Engineering Research Council of Canada (NSERC).
BMP is supported by an NSF Astronomy and Astrophysics Postdoctoral Fellowship under award AST2001663 JMC, JF, and JLI are supported by NSF/AST grant 2009894.
GB acknowledges support from: the Agencia Estatal de Investigaci{\'o}n del Ministerio de Ciencia en Innovaci{\'o}n (AEI-MICIN) and the European Regional Development Fund (ERDF) under grants
{\tiny CEX2019-000920-S and PID2020-118778GB-I00/10.13039/501100011033}.
\end{acknowledgments}

\software{{\tt Sextractor} \citep{sextra}, {\sc IRAF} \citep{iraf93}, Topcat \citep{topcat05}, SuperMongo \citep{sm91}, {\tt R}, {\tt JAGS}.}
\facilities{VLT:Yepun(MUSE)}


\end{document}